\documentclass[12pt,preprint]{aastex63}
\usepackage{xcolor}
\usepackage{graphicx}
\usepackage{hyperref}

\usepackage[caption=false]{subfig}
\usepackage{amsmath}

\begin{document}

\title{Flux Calibration of CHIME/FRB Intensity Data}
\correspondingauthor{Bridget C. Andersen}
\email{bridget.andersen@mail.mcgill.ca}
\author[0000-0001-5908-3152]{Bridget C.~Andersen}
  \affiliation{Department of Physics, McGill University, 3600 rue University, Montr\'eal, QC H3A 2T8, Canada}
  \affiliation{Trottier Space Institute, McGill University, 3550 rue University, Montr\'eal, QC H3A 2A7, Canada}
\author[0000-0003-3367-1073]{Chitrang Patel}
  \affiliation{Department of Physics, McGill University, 3600 rue University, Montr\'eal, QC H3A 2T8, Canada}
  \affiliation{Trottier Space Institute, McGill University, 3550 rue University, Montr\'eal, QC H3A 2A7, Canada}
\author[0000-0002-1800-8233]{Charanjot Brar}
  \affiliation{Department of Physics, McGill University, 3600 rue University, Montr\'eal, QC H3A 2T8, Canada}
  \affiliation{Trottier Space Institute, McGill University, 3550 rue University, Montr\'eal, QC H3A 2A7, Canada}
\author[0000-0001-8537-9299]{P.~J.~Boyle}
  \affiliation{Department of Physics, McGill University, 3600 rue University, Montr\'eal, QC H3A 2T8, Canada}
  \affiliation{Trottier Space Institute, McGill University, 3550 rue University, Montr\'eal, QC H3A 2A7, Canada}
\author[0000-0001-8384-5049]{Emmanuel Fonseca}
  \affiliation{Department of Physics and Astronomy, West Virginia University, P.O. Box 6315, Morgantown, WV 26506, USA }
  \affiliation{Center for Gravitational Waves and Cosmology, West Virginia University, Chestnut Ridge Research Building, Morgantown, WV 26505, USA}
\author[0000-0001-9345-0307]{Victoria M.~Kaspi}
  \affiliation{Department of Physics, McGill University, 3600 rue University, Montr\'eal, QC H3A 2T8, Canada}
  \affiliation{Trottier Space Institute, McGill University, 3550 rue University, Montr\'eal, QC H3A 2A7, Canada}
\author[0000-0002-4279-6946]{Kiyoshi W.~Masui}
  \affiliation{MIT Kavli Institute for Astrophysics and Space Research, Massachusetts Institute of Technology, 77 Massachusetts Ave, Cambridge, MA 02139, USA}
  \affiliation{Department of Physics, Massachusetts Institute of Technology, 77 Massachusetts Ave, Cambridge, MA 02139, USA}
\author[0000-0002-0772-9326]{Juan Mena-Parra}
  \affiliation{Dunlap Institute for Astronomy \& Astrophysics, University of Toronto, 50 St.~George Street, Toronto, ON M5S 3H4, Canada}
  \affiliation{David A.~Dunlap Department of Astronomy \& Astrophysics, University of Toronto, 50 St.~George Street, Toronto, ON M5S 3H4, Canada}
\author[0000-0003-2095-0380]{Marcus Merryfield}
  \affiliation{Department of Physics, McGill University, 3600 rue University, Montr\'eal, QC H3A 2T8, Canada}
  \affiliation{Trottier Space Institute, McGill University, 3550 rue University, Montr\'eal, QC H3A 2A7, Canada}
\author[0000-0001-8845-1225]{Bradley W.~Meyers}
  \affiliation{International Centre for Radio Astronomy Research (ICRAR), Curtin University, Bentley WA 6102 Australia}
  \affiliation{Department of Physics and Astronomy, University of British Columbia, 6224 Agricultural Road, Vancouver, BC V6T 1Z1 Canada}
\author[0000-0003-3154-3676]{Ketan R.~Sand}
  \affiliation{Department of Physics, McGill University, 3600 rue University, Montr\'eal, QC H3A 2T8, Canada}
  \affiliation{Trottier Space Institute, McGill University, 3550 rue University, Montr\'eal, QC H3A 2A7, Canada}
\author[0000-0002-7374-7119]{Paul Scholz}
  \affiliation{Dunlap Institute for Astronomy \& Astrophysics, University of Toronto, 50 St.~George Street, Toronto, ON M5S 3H4, Canada}
\author[0000-0003-2631-6217]{Seth R.~Siegel}
  \affiliation{Department of Physics, McGill University, 3600 rue University, Montr\'eal, QC H3A 2T8, Canada}
  \affiliation{Trottier Space Institute, McGill University, 3550 rue University, Montr\'eal, QC H3A 2A7, Canada}
  \affiliation{Perimeter Institute for Theoretical Physics, 31 Caroline Street N, Waterloo, ON N25 2YL, Canada}
\author[0000-0001-7755-902X]{Saurabh Singh}
  \affiliation{Raman Research Institute, Sadashivanagar, Bengaluru, India}
\newcommand{\allacks}{
B.\,C.\,A. is supported by an FRQNT Doctoral Research Award.
K.\,W.\,M. holds the Adam J. Burgasser Chair in Astrophysics and is supported by an NSF Grant (2008031).
M.\,Me. is supported by an NSERC PGS-D award.
P.\,S. is a Dunlap Fellow.
V.\,M.\,K. holds the Lorne Trottier Chair in Astrophysics \& Cosmology, a Distinguished James McGill Professorship, and receives support from an NSERC Discovery grant (RGPIN 228738-13), from an R. Howard Webster Foundation Fellowship from CIFAR, and from the FRQNT CRAQ.
}

\shorttitle{}
\shortauthors{Andersen et al.}
\submitjournal{\aj}


\begin{abstract}
    Fast radio bursts (FRBs) are bright radio transients of micro-to-millisecond duration and unknown extragalactic origin. Central to the mystery of FRBs are their extremely high characteristic energies, which surpass the typical energies of other radio transients of similar duration, like Galactic pulsar and magnetar bursts, by orders of magnitude. Calibration of FRB-detecting telescopes for burst flux and fluence determination is crucial for FRB science, as these measurements enable studies of the FRB energy and brightness distribution in comparison to progenitor theories. The Canadian Hydrogen Intensity Mapping Experiment (CHIME) is a radio interferometer of cylindrical design. This design leads to a high FRB detection rate but also leads to challenges for CHIME/FRB flux calibration. This paper presents a comprehensive review of these challenges, as well as the automated flux calibration software pipeline that was developed to calibrate bursts detected in the first CHIME/FRB catalog, consisting of 536 events detected between July 25th, 2018 and July 1st, 2019. We emphasize that, due to limitations in the localization of CHIME/FRB bursts, flux and fluence measurements produced by this pipeline are best interpreted as lower limits, with uncertainties on the limiting value.
\end{abstract}

\section{Introduction}
Fast radio bursts (FRBs) are an enigmatic class of bright microsecond- to millisecond-duration radio transients of extragalactic origin. Over the past decade since their discovery \citep{lbm+07}, more than six hundred bursts have been detected across surveys of varying capabilities. These detections have driven significant progress in the observational characterization of FRBs, revealing a sprawling landscape of diverse population properties \citep{petroff_review_2022}. Despite this observational progress, as of yet, a comprehensive physical explanation of the FRB phenomenon remains elusive.

One of the key FRB properties that theories must contend with is their extreme energetics. Both repeating and so-far non-repeating FRBs have been localized to cosmological distances ranging approximately from $150$\,Mpc to $4$\,Gpc which, paired with well-constrained fluence measurements, imply isotropic-equivalent burst energies spanning at least six orders of magnitude from $10^{36}$ to $10^{42}$\,erg (\citealt{mnh+20}; \citealt{rcd+19}; energies here calculated assuming a fiducial burst bandwidth of $500$\,MHz). Even within the population of bursts detected from a single repeating source the energy range can be vast. For example, burst energies from FRB 121102 have ranged from $10^{37}$ to $10^{40}$\,erg (\citealt{gms+19}; \citealt{ola+17}). The FRB energetics problem is thus multifaceted: progenitor and emission models must account not only for such extreme energy outputs within short burst durations, but also for a large range of energy outputs both across the entire population and within a single repeating source. Beyond minimum and maximum energy limits, FRB brightness serves as a valuable metric for many other tests of FRB origins, including constraints on the cumulative fluence distribution ($\log N{-}\log S$; e.g., \citealt{me18a}, \citealt{catalog1}), the underlying energy and distance distribution (e.g., \citealt{james2022}, \citealt{shin23}), individual repeater energy distributions (e.g., \citealt{lab+17}, \citealt{gms+19}, \citealt{css+21}), and the dispersion$-$brightness relation \citep{smb+18}. Additionally, fluence measurements are crucial for FRB rate determination \citep{catalog1} and observational follow-up strategies.

The revelatory potential of these comparisons motivates accurate physical measurements of FRB flux densities and fluences. The typical method for radio transient flux calibration is the radiometer equation, which appears in various forms throughout pulsar and FRB astronomy. Derived from physical first principles, the radiometer equation provides a process for conversion between instrumental units and physical flux density using just a few fundamental telescope parameters,
\begin{equation} \label{eq:radiometer}
    S_{\text{det}} = \frac{(S/N) \; T_{\text{sys}}}{G \sqrt{n_{\text{p}} t_{\text{s}} \Delta f}} = \frac{(S/N) \; \text{SEFD}}{\sqrt{n_{\text{p}} t_{\text{s}} \Delta f}},
\end{equation}
where $S_{\text{det}}$ is the detected flux density, $S/N$ is the signal-to-noise of the burst time series, $T_{\text{sys}}$ is the system temperature, $G$ is the telescope gain, $n_p$ is the number of summed polarizations, $t_{\text{s}}$ is the sampling time of the telescope, $\Delta f$ is the bandwidth of each frequency channel, and SEFD is the system equivalent flux density (see, e.g., \citealt{lk04}). The power of this calibration method is its simplicity, since it requires knowledge of just a few system parameters: either $T_{\text{sys}}$ and $G$ or the SEFD. Given its robustness, the radiometer equation is used to calibrate data in nearly all FRB experiments and surveys in operation today (e.g., Arecibo Pulsar ALFA Survey, \citealt{sch+14}; Parkes High Time Resolution Universe Survey, \citealt{cpk+16}; Green Bank Northern Celestial Cap Pulsar Survey, \citealt{pck+20}; Commensal Radio Astronomy Five-hundred-meter Aperture Spherical radio Telescope Survey or FAST, \citealt{zll+20}; Upgraded Molonglo Observatory Synthesis Telescope, \citealt{ffb+19}). Its use in the field is so routine that the flux calibration descriptions in many FRB papers consist of a short paragraph or table listing the relevant system parameters.

Values for $T_{\text{sys}}$, $G$, or SEFD are typically determined using sources of known intensity, such as a bright astrophysical source of radio emission (e.g., the Australian Square Kilometre Array Pathfinder, ASKAP, observes a Seyfert galaxy of well-known flux, PKS B1934-638; \citealt{mab+16}), a previously calibrated noise diode (e.g., FAST calibrates through injection of a $T \sim 11$\,K noise diode; \citealt{jth+20}), or a pair of hot and cold ``loads'' of known temperature (e.g., a radio-absorbing material in an oven or liquid nitrogen bath, as used to calibrate the Parkes Ultra-Wide Bandwidth Receiver; \citealt{hmd+20}). Once these values are measured, they can change over time due to variations in the telescope structure, receiver electronics, or environmental conditions. Based on their system design, FRB experiments make different assumptions about the temporal stability of the measured system parameters. For example, the gain values quoted in current FRB papers for single-dish telescopes like the Parkes Multibeam receiver and the Green Bank Telescope (GBT) were measured years ago, in 1996 \citep{swb+96} and 2009\footnote{\url{https://www.gb.nrao.edu/~rmaddale/GBT/ReceiverPerformance/PlaningObservations.htm}}, respectively. These systems have cryogenically cooled receivers, which tend to be stable over long periods of time. In contrast, newer non-cooled interferometric experiments like ASKAP rely on calibrations from within several months to within several hours of an observed FRB (e.g., ASKAP's fly's eye mode observations are calibrated within months to days: \citealt{bsm+17}, \citealt{smb+18}, whereas interferometric observations are calibrated within hours: \citealt{bdp+19}).

Beyond the determination of system parameters, the most central obstacle to accurate FRB flux calibration is burst localization. Unlike pulsars, most FRBs observationally appear to burst only once. This presents a particularly difficult challenge, as FRB-detecting telescopes must often rely on just a single burst for localization purposes. Most telescopes in FRB detection history have been limited in their ability to spatially sample their field of view, leading to large uncertainties in detected burst positions within the telescope beam pattern (e.g., for GBT- and Parkes-detected FRBs, burst position is reported as the center of the beam pointing, with an uncertainty corresponding to the beam FWHM; \citealt{ssh+16b}, \citealt{cpk+16}). Without precise knowledge of the location of the FRB in the beam, measured fluences cannot be properly corrected for attenuation due to the beam response. As a result, the vast majority of FRB fluences published to-date are lower limits, calculated under the simplifying assumption that the FRBs were detected at beam boresight. Exceptions include fluences determined for repeating FRBs of known position \citep[e.g.,][]{lwz+21}, or one-off FRBs detected using long-baseline interferometric telescopes with sub-arcsecond localization capabilities (e.g., ASKAP, \citealt{bsm+17}; the Deep Synoptic Array-110 or DSA-110, \citealt{rcc+23}).

The Canadian Hydrogen Intensity Mapping Experiment (CHIME) is a radio interferometer located near Penticton, B.C., that has detected an unprecedented number of FRBs \citep{catalog1}. The efficiency with which CHIME detects these bursts is enabled by its novel design, consisting of four $20$-m by $100$-m cylindrical reflectors with $256$ dual-polarization feeds lined along each axis that are sensitive to a wide bandwidth of $400{-}800$\,MHz \citep{chime_overview}. The CHIME Fast Radio Burst (CHIME/FRB) project operates commensally on the CHIME data stream, continuously searching total intensity data from a grid of $1{,}024$ formed beams over a ${\sim}200$ square degree field of view at $1$-ms time resolution \citep{chimefrb_overview}. Within its first year of operation from July 25th, 2018 to July 1st, 2019, CHIME/FRB detected a catalog of $536$ new FRBs \citep{catalog1}, leading to important discoveries including confirmation of the existence of FRB emission down to $400$\,MHz \citep{abb+19a}, the discovery of $47$ new repeating FRB sources \citep{abb+19b,abb+19c,fab+20,bgk+21,aac+22,mckinven22,abb+23}, the discovery of periodicity in the active window of a repeating FRB \citep{aab+20}, the discovery of sub-second periodicity within a single FRB burst profile \citep{abb+22}, the detection of an FRB in the nearby galaxy M81 \citep{bgk+21}, and the detection of a bright FRB-like burst from a Galactic magnetar \citep{abb+20}.
 
While the CHIME's novel design is particularly effective for FRB detection, it also introduces novel challenges for flux calibrating bursts detected in CHIME/FRB intensity data. In this paper, we provide a comprehensive review of these challenges (Section~\ref{sec:calibration_challenges}) and we describe an automated flux calibration software pipeline (Section~\ref{sec:automated_calibration_pipeline}) that was developed to calibrate bursts detected in the first CHIME/FRB catalog \citep{catalog1}.

\section{CHIME/FRB Calibration Challenges} \label{sec:calibration_challenges}

The requirements for accurate FRB flux calibration distill into three main components: determination of the time-variable system sensitivity, precise FRB localization, and characterization of the telescope beam response as a function of frequency and on-sky location. 

For the first bursts detected from CHIME/FRB, this flux calibration process became an especially elaborate challenge requiring the characterization of a novel and still-developing system within an accelerated timeframe. By July 20th, 2018, less than a year after first light for the CHIME telescope on September 7th, 2017, a limited version of the CHIME/FRB beamforming and detection pipeline was running semi-stably. On July 25th, less than a week later, the first FRB was detected \citep{abb+19a}. At this point, the system was still in a pre-commissioning period, where only a small and variable number of beams were being searched at any given time and the complex gain calibration strategy was under development and constantly changing. Despite the instability of the system during this time, $14$ FRBs were detected before pre-commissioning ended and the full $1{,}024$ beam system was implemented on August 27th, 2018 \citep{abb+19a}. Since then, CHIME/FRB has steadily detected FRBs, and $536$ FRBs were accumulated within the first year of operation \citep{catalog1}. 

These frequent detections early in the commissioning of the experiment, and ensuing scientific results, necessitated the development of the CHIME/FRB flux calibration pipeline before many aspects of the system were fully operational, let alone thoroughly quantified or understood. Here we highlight the state of our knowledge of the system during the time period encompassing the first catalog (July 25th, 2018 to July 1st, 2019) in order to give context for the design and limitations of the automated pipeline presented in Section~\ref{sec:automated_calibration_pipeline}.

\subsection{Characterization of System Parameters} \label{sec:characterization_sys_params} 
Robust measurements of key system parameters for CHIME/FRB had not yet been completed at the start of flux calibration pipeline development, although a few nominal values existed. Based on design specifications, the receiver temperature for the full analog chain of CHIME was expected to be approximately $50$\,K \citep[including spillover noise from the ground;][]{baa+14}. The gain at peak sensitivity could also be very roughly determined by estimating the illuminated area of the reflectors, yielding a value of $G_0 \sim 1.2$\,K\,Jy$^{-1}$. Later, in May and June of 2019, members of the CHIME/Cosmology collaboration used a pair of hot and cold loads to measure the receiver (${\sim}20{-}25$\,K, as a function of frequency) and system (${\sim}50{-}60$\,K, as a function of frequency) temperatures, confirming the design specifications \citep{chime_overview}. 
    
\subsection{Complex Beam Pattern} \label{sec:complex_beam_pattern} 

The motivating tenet of the CHIME/FRB beamforming scheme is to spatially sample the CHIME field of view at the highest sensitivity and frequency resolution possible within the computational budget. To accomplish this goal, CHIME/FRB forms a closely packed grid of $1{,}024$ static beams for each polarization and frequency. Arranged in four columns East-West and $256$ rows North-South, the synthesized beams tile the ${\sim}200$ square degree span of CHIME's primary beam, allowing for real-time high-sensitivity FRB detection and localization across a large instantaneous swath of sky.

Forming so many simultaneous beams is a computationally intensive process. To facilitate this process, CHIME/FRB uses an algorithm called Fast Fourier Transform (FFT; \citealt{nvp+17,msn+19}) beamforming to form beams in the North-South direction. This algorithm leverages the regular grid-like layout of CHIME's feeds to relax the $\mathcal{O}(N^2)$ runtime of conventional beamforming to $\mathcal{O}(N\log N)$, where $N$ is the number of feeds \citep[e.g.,][]{tz09,tz10}. However, for all of its computational advantages, FFT beamforming also significantly complicates the CHIME/FRB bandpass response. In particular, the FFT-formed beams (hereafter FFT beams) exhibit complex structure as a function of frequency, which manifests as sharp discontinuous and periodic spectral features that change shape significantly over small displacements on-sky. As a further complication, these FFT beamforming structures are superimposed over bandpass ripples in the primary beam response. Differentiating these spectral features from the intrinsic FRB spectra is a nontrivial task, which stands as one of the most fundamental challenges to CHIME/FRB flux calibration. In this section we review CHIME/FRB's synthesized beam response (Section~\ref{sec:fftbeamforming}), CHIME's primary beam response (Section~\ref{sec:primarybeam}), and the development of the CHIME/FRB beam model (Section~\ref{sec:beammodel}), to contextualize decisions made in the design of the flux calibration pipeline.

\subsubsection{FFT-Formed Beams} \label{sec:fftbeamforming}

CHIME/FRB operates in phased array mode by coherently summing feed signals with different time delays to constructively interfere in a particular direction. In brief, a composite beamformed signal at an observing frequency, $f$, directed at steering angle $\theta_m$, can be written as
\begin{equation} \label{eq:bf_fourier_domain}
    B(f,\theta_m) = \sum_{n=1}^{N} a_n \, X_n{\left[f\right]} \, e^{-i2\pi f \tau_n (\theta_m)}
\end{equation}
\noindent where $N$ is the number of feeds, $X_n{[f]}$ are the channelized complex data, $a_n$ are the constant gains applied to correct for instrumental delays, and $\tau_n (\theta_m)$ are the time delays required to point the beam to a specified direction \citep{mucci84}. In this formulation, the intensity data output from the beam is taken to be $|B(f,\theta_m)|^2$ \citep{maranda89}. Notice that forming $N$ beams using this method would take $\mathcal{O}(N^2)$ time.

The crux of FFT beamforming comes with the realization that the time delays required for beamforming an array of linearly spaced feeds with separation $d$ are given by: $\tau_n (\theta_m) = n \frac{d}{c} \sin (\theta_m)$, where $n$ is the feed index and $c$ is the speed of light. If we choose to form beams at steering angles
\begin{equation} \label{eq:bf_fft_angles}
    \theta_m = \sin^{-1} \left(\frac{c\,m}{f\,N\,d}\right)
\end{equation}
\noindent then, by substituting $\tau_n (\theta_m)$, Equation~\ref{eq:bf_fourier_domain} becomes:
\begin{equation} \label{eq:bf_fft}
    B(f,\theta_m) = \sum_{n=1}^{N} a_n\,X_n{\left[f\right]}\,e^{-i2\pi m n / N}
\end{equation}
This is just a discrete Fourier transform, mapping the spatial offsets in feed positioning to angular beams on the sky. This expression can be evaluated to form $N$ beams using a fast Fourier transform in $\mathcal{O}(N\log N)$ time. 

\begin{figure}[ht!]
    \centering
    \includegraphics[width=1.\columnwidth]{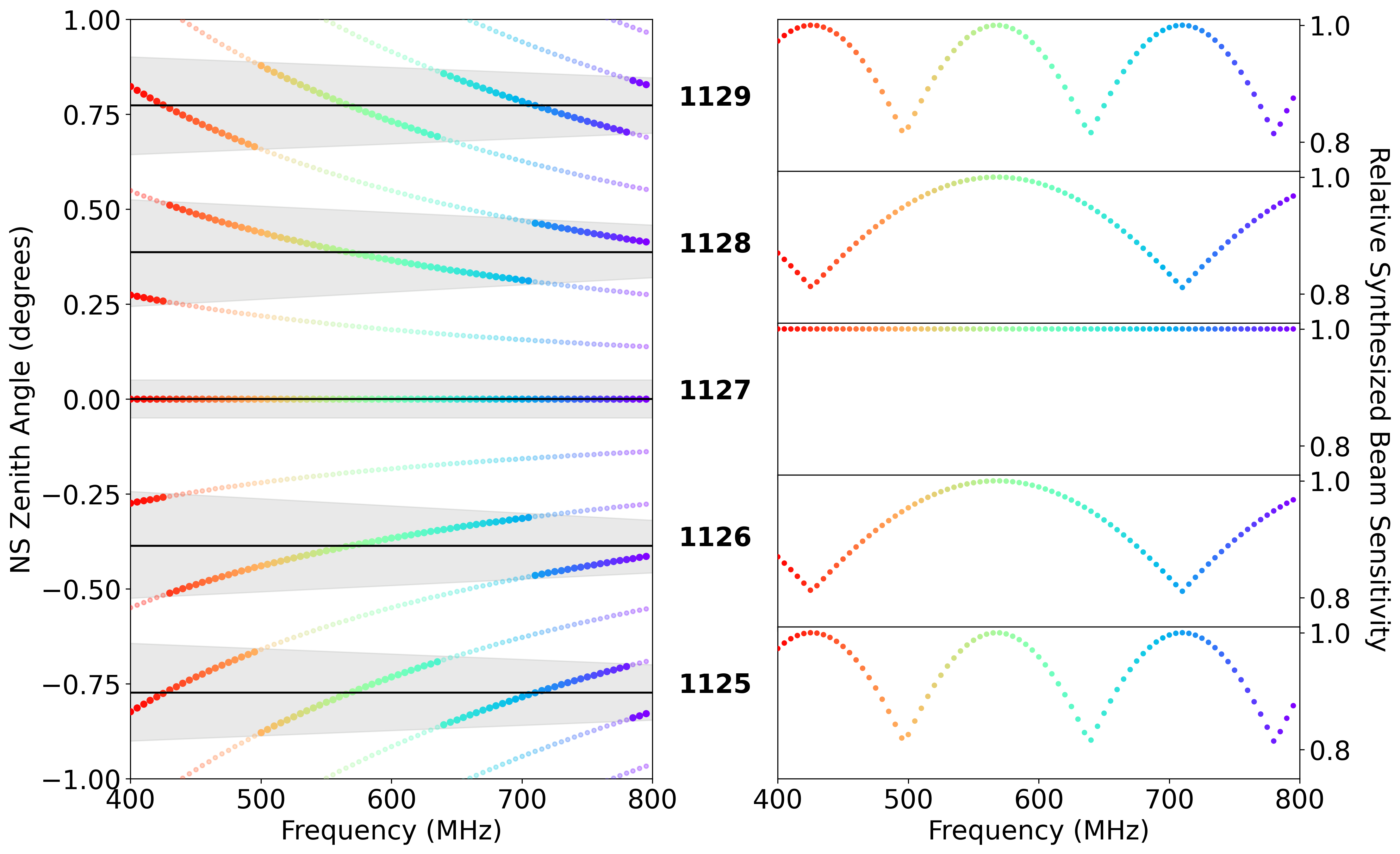}
    \caption{Illustration of the chromatic effect of FFT beamforming and clamping across the CHIME/FRB bandwidth. This figure expands upon Figure~1 from \cite{nvp+17}. (Left) The position of the dots represents the location of nominal peak sensitivity for each formed beam at a particular frequency. The colored dots enclosed by the gray areas are the selected nearest-neighbor clamped beams, whereas the fainter colored dots are the additional discarded beams. Horizontal black lines represent the nominal clamped beam centers, labeled by the corresponding beam number. (Right) The sensitivity versus frequency at the center of each of the clamped beams in the left panel.} \label{fig:clamping_and_sensitivities}
\end{figure}

\begin{figure}[ht!]
  \centering
  \includegraphics[width=0.46\linewidth]{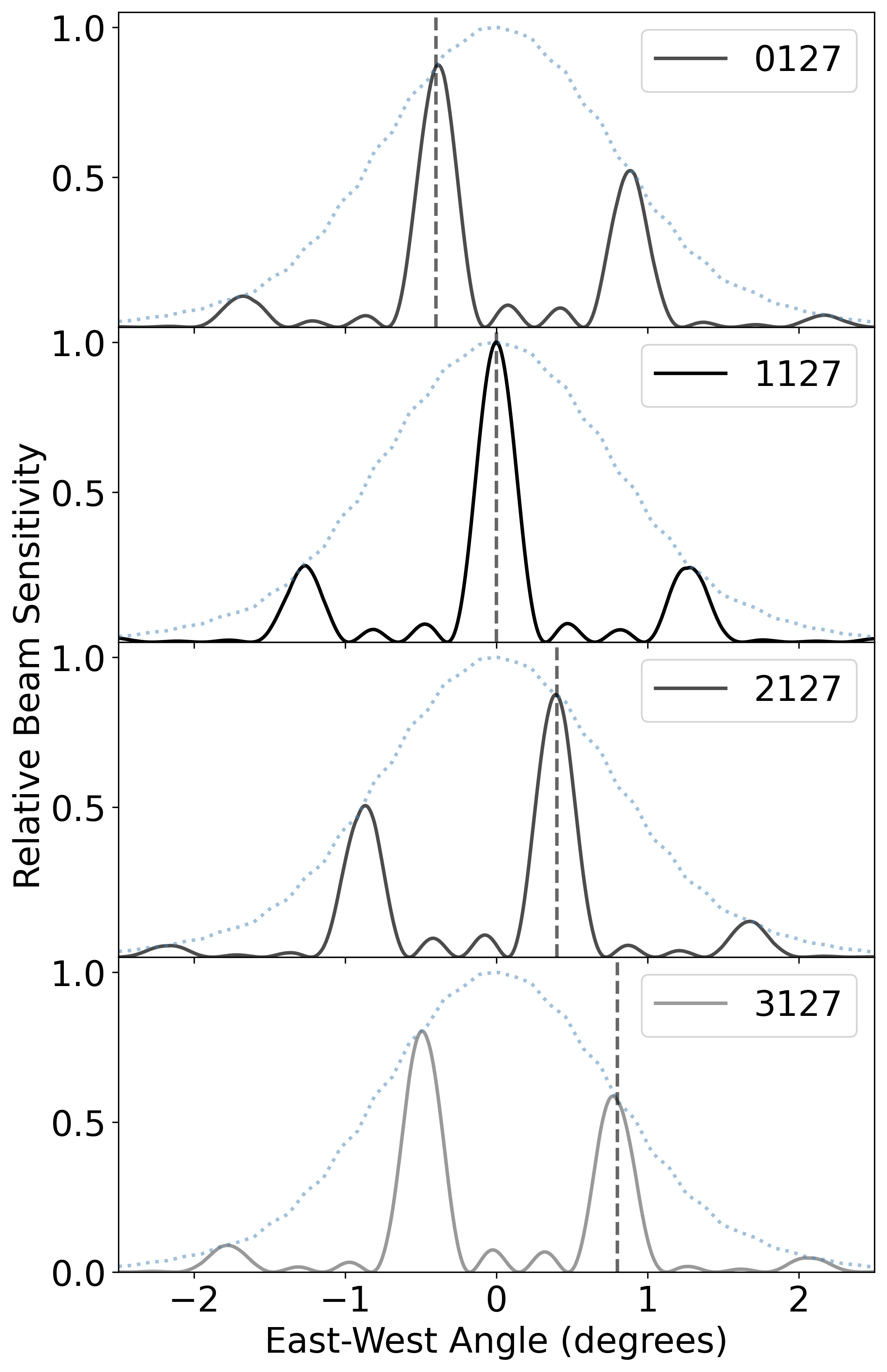} %
  \includegraphics[width=0.46\linewidth]{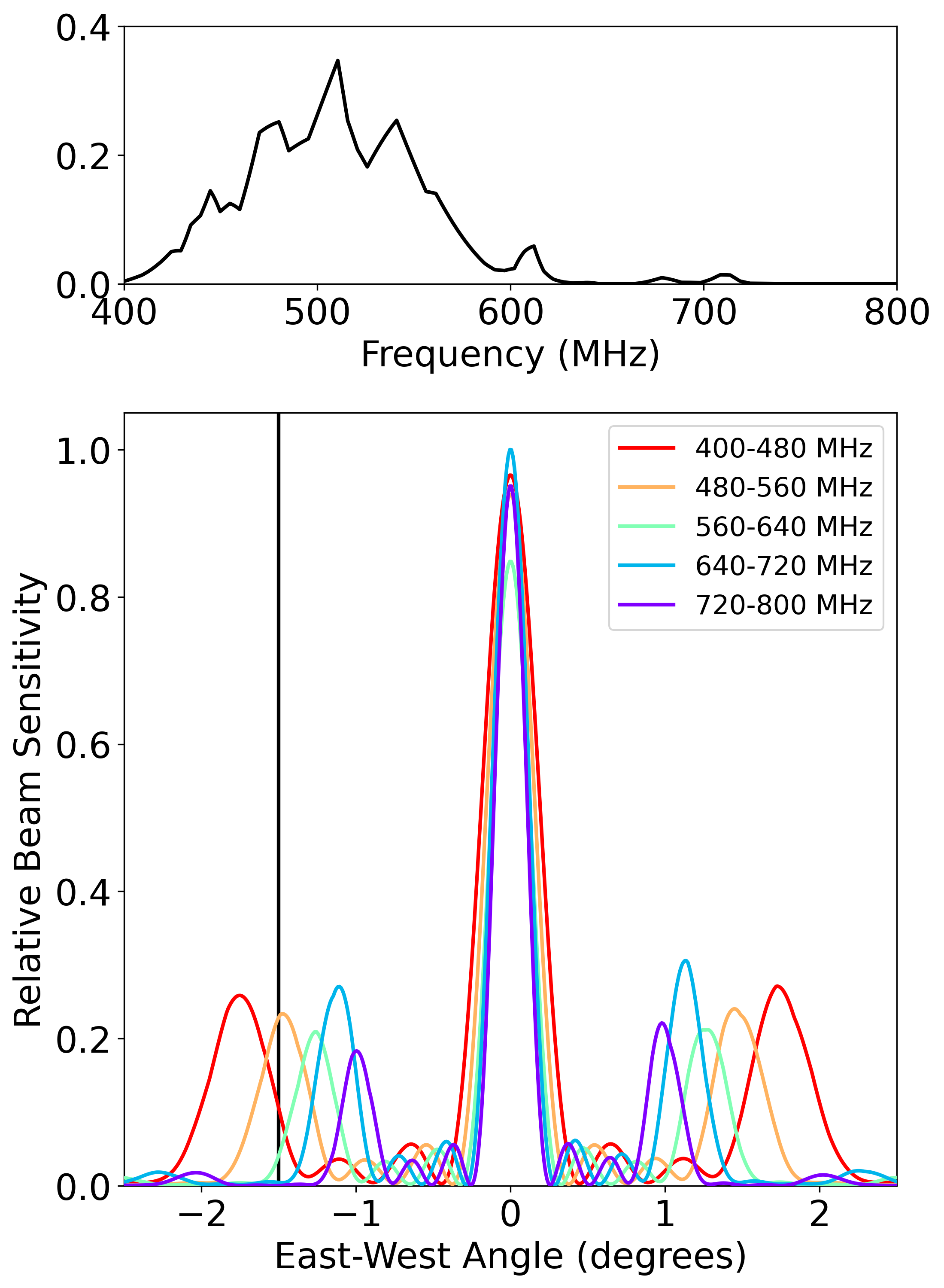}
  \caption{Representations of the East-West sensitivity profile of the CHIME/FRB formed beams. (Left) A plot of the East-West profile for the row of beams at zenith at $600$\,MHz (solid lines), which are attenuated by the primary beam envelope (dotted blue lines). The dashed vertical lines indicate the main lobe of each beam. (Right) The East-West profile of beam $1127$. (Bottom) The profile is split into five subbands, with colored lines representing each subband (red corresponding to low frequencies and purple corresponding to high frequencies). (Top) The frequency response of the beam at the sidelobe location marked by a vertical black line in the bottom plot.}\label{fig:east_west_profile}
\end{figure}

\begin{figure}[ht!]
  \centering
  \includegraphics[width=.475\linewidth]{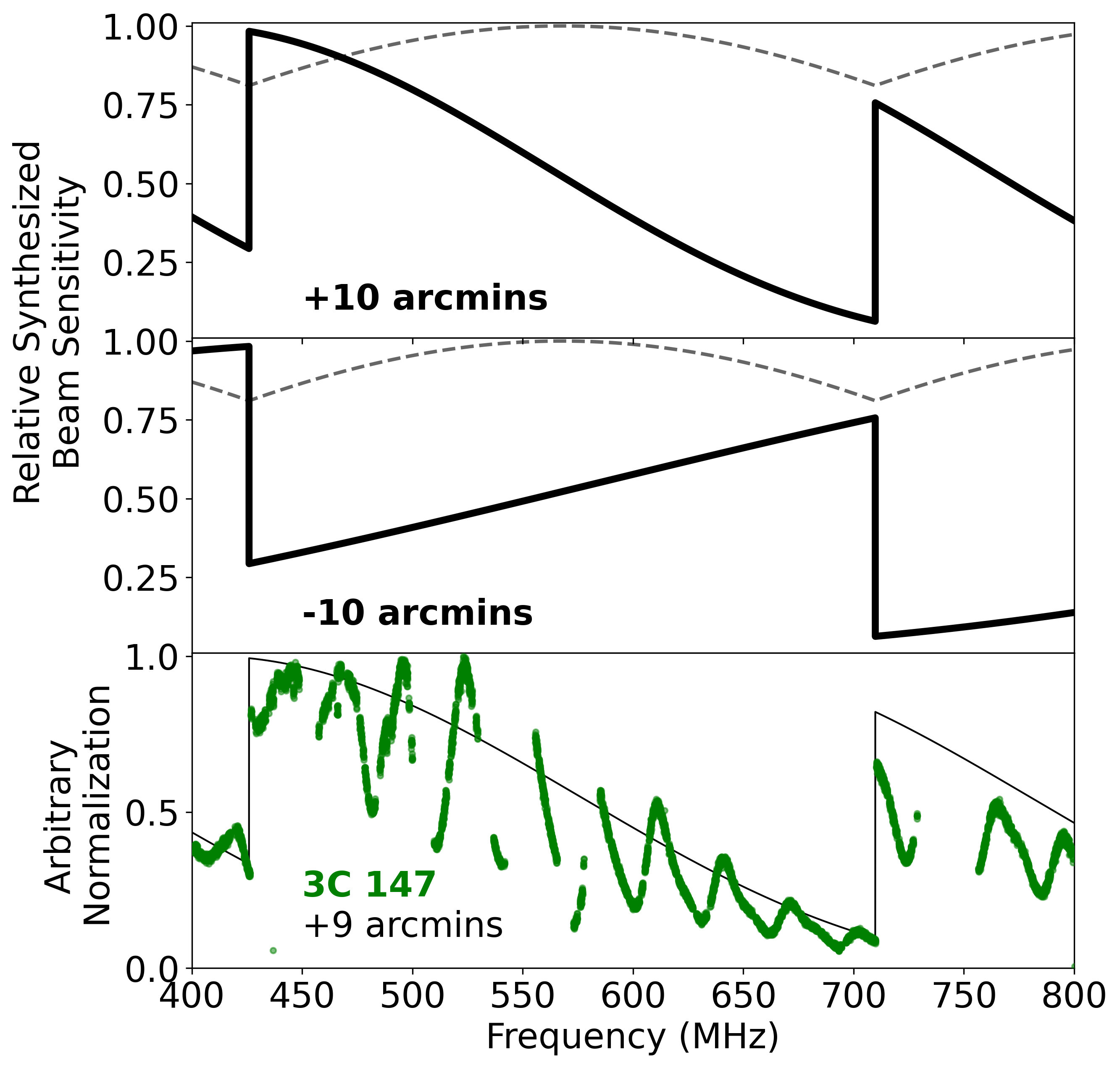}
  \caption{Variations in the response of synthesized beam $1128$ as a function of frequency and location within the beam. The thick black lines represent the sensitivity versus frequency $10$ arcminutes north of the center of beam $1128$ (Top) and $10$ arcminutes south (Middle). The thin grey dashed line in each of these plots represents the response at beam center. The bottom plot shows the synthesized beam response at the transit location of radio galaxy 3C 147 (thick black line), along with a background-subtracted and normalized observed spectrum of the source with CHIME/FRB (green points). Note that 3C 147 is located $9$ arcminutes north of the center of beam $1128$.}\label{fig:in_beam_clamping}
\end{figure}

\begin{figure}[ht!]
    \centering
    \includegraphics[width=1.\columnwidth]{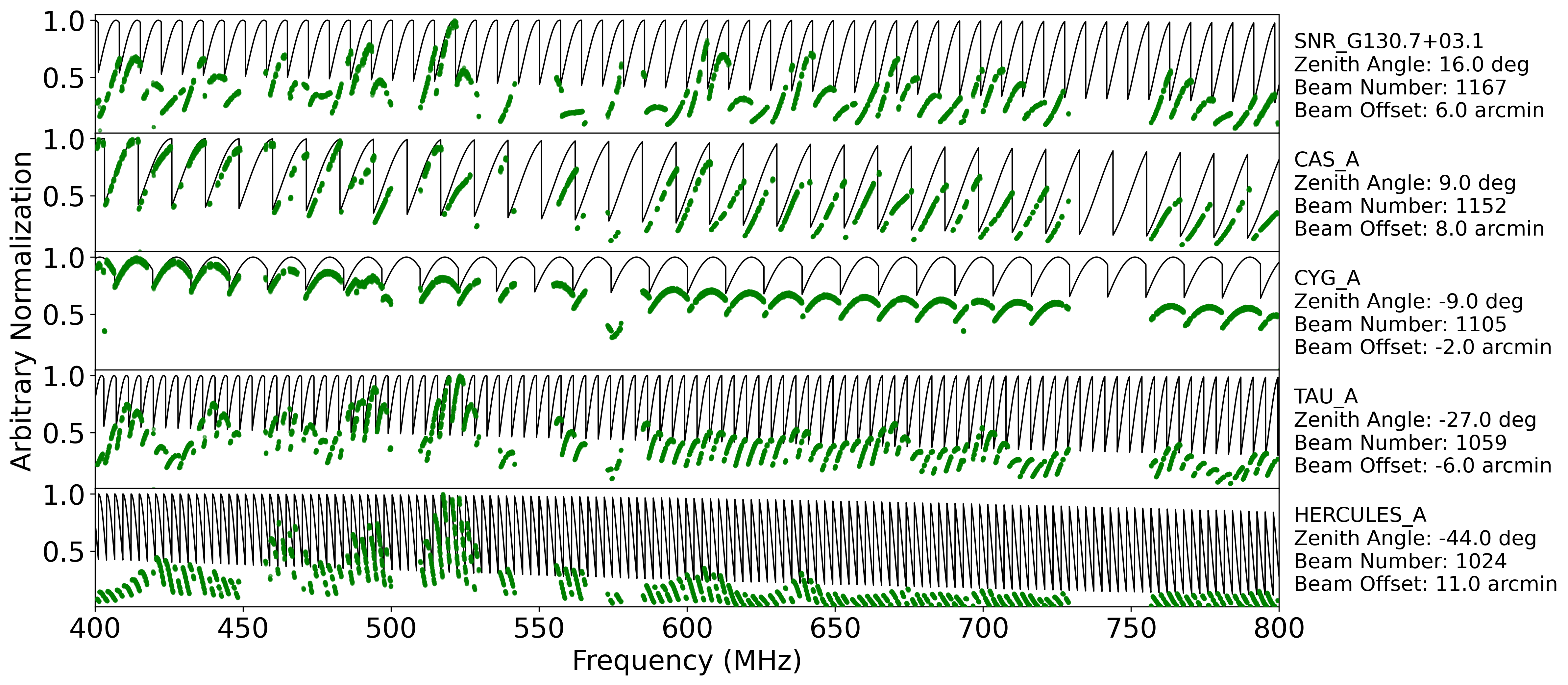}
    \caption{A sampling of CHIME/FRB synthesized beam responses at the locations of several bright supernova remnant and radio galaxy transits (black lines), along with background-subtracted and normalized observed spectra of each source (green points). The source name, zenith angle, beam number, and offset from beam center are labeled to the right of each plot.} \label{fig:clamping_samples}
\end{figure}

FRB surveys typically need to maximize broadband sensitivity to a single sky location to increase detection significance and allow for FRB spectral characterization. However, as dictated by Equation~\ref{eq:bf_fft_angles}, the steering angles of FFT-formed beams are dependent on frequency as $\sin (\theta_m) \propto 1 / f$, causing formed beams at higher frequencies to be closer together than those at lower frequencies. This effect chromatically smears the sensitivity pattern of a single beam (indexed by $m$) across the sky. To reduce this effect with CHIME/FRB, we use a method called ``nearest-neighbor clamping'' \citep{maranda89,nvp+17}. First, the $256$ North-South feed inputs are zero-padded by a factor of two so that the FFT in Equation~\ref{eq:bf_fft} forms $512$ closely packed beams by Fourier interpolation. These beams are then subsampled (or ``clamped'') to form $256$ beams at the desired pointings. The most sensitive beam for each frequency is chosen for each pointing, forming a ``Frankenstein'' beam of combined components. This process is illustrated in the left panel of Figure~\ref{fig:clamping_and_sensitivities} for a few CHIME/FRB beams near zenith, where the steering angle is North-South zenith angle.\footnote{Note that the code for producing Figures~\ref{fig:clamping_and_sensitivities}-\ref{fig:clamping_samples} is publicly available at \href{https://github.com/bandersen441/chimefrb_intensity_flux_calibration_plots}{on Github}.}

The full CHIME/FRB beamforming pipeline is a hybrid of two techniques: FFT beamforming (Equation~\ref{eq:bf_fft}) with clamping is used to form the $256$ rows of beams in the North-South direction, while brute-force phasing (Equation~\ref{eq:bf_fourier_domain}) is used to form the four columns of beams in the East-West direction. For the majority of the CHIME/FRB experiment duration, the configuration of the beam grid has been fixed. In the North-South direction, the beams are equally spaced in $\sin{\theta}$, where $\theta$ is the zenith angle. One beam is centered at zenith ($\theta = 0^{\circ}$), while $127$ beams tile to $\theta = -60^{\circ}$ South and $128$ beams tile to $\theta = 60^{\circ}$ North. Beams get more elongated North-South at larger zenith angles, because the projected baselines between feeds shorten nearer to the horizon. In the East-West direction, the configuration is asymmetrical: one column is centered along the meridian while another column is formed $0.4^{\circ}$ to the East and two other columns are formed $0.4^{\circ}$ and $0.8^{\circ}$ to the West (see the left panel in Figure~\ref{fig:east_west_profile}). The resulting formed beams are labeled with integers according to their location in the grid: the rows South to North have an index ranging from $0$ to $255$ and the East to West columns add a factor of $0$, $1000$, $2000$, or $3000$ to the resulting index. For example, the beam at zenith in the meridian column corresponds to $127 + 1000 = 1127$.

The full implications of this hybrid beamforming scheme are complex and more effectively shown than told. Aided by figures, in the remainder of this section and the next we highlight some more detailed but central aspects of CHIME/FRB's beam response that influence the design of the flux calibration pipeline.

\paragraph{Severe Variations in North-South Spectral Structure}
As illustrated in Figure~\ref{fig:clamping_and_sensitivities}, the clamping algorithm used by CHIME/FRB introduces periodic cusp-like structures in the resulting sensitivity versus frequency due to chromatic spatial smearing within the clamped beam extent. These cusp-like features, called ``clamps,'' change shape and severity depending on the beam being considered as well as the location on-sky. The right panel of Figure~\ref{fig:clamping_and_sensitivities} shows the beam response versus frequency at the center of five clamped beams surrounding zenith. Notably, the number of clamps in the CHIME/FRB bandwidth increases with the zenith angle of the clamped beam, which can cause a particularly complex response pattern for beams closer to the horizon. For example, beam $1128$ at $0.4^{\circ}$ from zenith has only two clamps, while beam $1024$ at $-44^{\circ}$ zenith angle has a total of $145$ clamps. 

The shape of the clamps also changes over spatial displacements within a single beam. In particular, as you move away from beam center, the cusp-like structures turn into more severe discontinuous jumps. Figure~\ref{fig:in_beam_clamping} demonstrates this by showing the response of beam $1128$ at the edges of its nominal ${\sim}20^{\prime}$ North-South full-width half-maximum (FWHM; at $600$\,MHz). The bottom panel shows a CHIME/FRB observed spectrum from the transit of radio galaxy 3C 147 across beam $1128$, which exhibits the sharp clamping discontinuities at ${\sim}430$ and ${\sim}710$\,MHz. Note that the remaining oscillation in the spectrum is the $30$\,MHz ripple from the primary beam response (see Section~\ref{sec:primarybeam}). 

Figure~\ref{fig:clamping_samples} shows a sampling of other CHIME/FRB synthesized beam responses corresponding with data from the transits of several bright supernova remnants and radio galaxies. This sampling covers a wide range of zenith angles, beam numbers, and offsets from the beam center, which gives a qualitative demonstration of the complexity of the North-South synthesized beam response.

\paragraph{Chromatic Sidelobes in East-West Beam Profile}
The East-West profile of the CHIME/FRB formed beams consists of an intrinsic profile governed by exact phasing, which is then further attenuated by the primary beam response. The exact phasing profile has significant sidelobes of increased sensitivity due to the periodic nature of the Fourier transform (Equation~\ref{eq:bf_fourier_domain}). These sidelobes remain significant even with attenuation from the primary beam, as shown in the left panel of Figure~\ref{fig:east_west_profile}. In particular, beams in the $3000$ column have a sidelobe that is more sensitive than the main lobe in the middle of the CHIME/FRB band. 

Another notable property of these sidelobes is their smeared frequency response. The sidelobes of a given beam spread chromatically East-West, as shown in the right panel of Figure~\ref{fig:east_west_profile}, with lower frequencies spreading further than higher frequencies. This means that a burst detected in the East-West sidelobe of a given synthesized beam may be completely attenuated to the noise floor at some frequencies, making it appear band-limited. This is demonstrated in the top right panel of Figure~\ref{fig:east_west_profile}.

\subsubsection{Primary Beam} \label{sec:primarybeam}

\begin{figure}[ht!]
    \centering
    \includegraphics[width=.5\columnwidth]{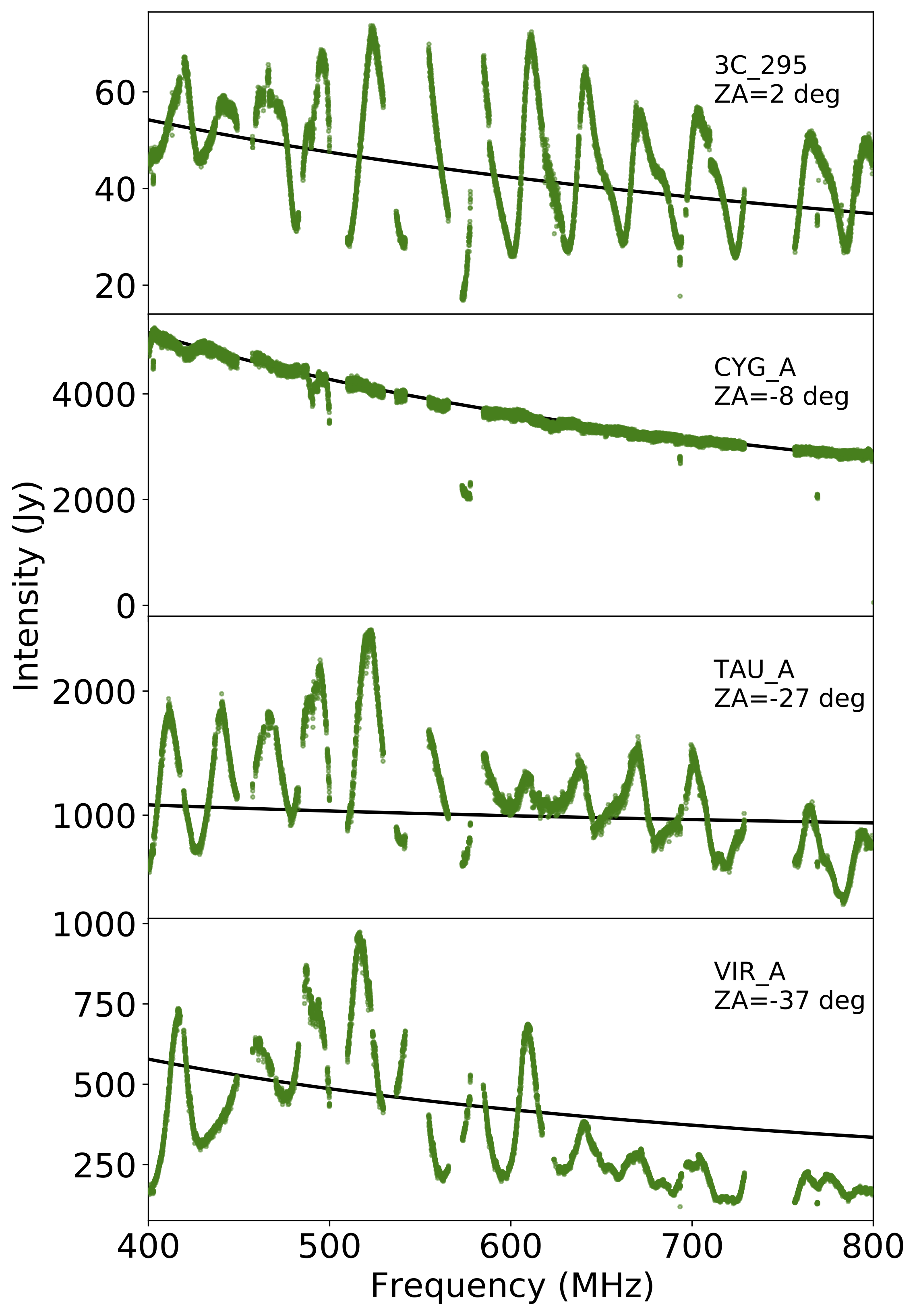}
    \caption{Representations of the sensitivity profile of the CHIME primary beam. The green points in each panel show the observed spectrum for a steady source at meridian transit that has been corrected for the synthesized beam response. The solid black lines show the actual known spectrum of each steady source. Each plot is labeled with the source name and zenith angle. Note that the spectrum for Cygnus A shows minimal variations from the primary beam ripple, as it was used as the complex gain calibrator at the time the data for this plot were taken (see Section~\ref{sec:primarybeam}).} \label{fig:primary_beam_profiles}
\end{figure}

The full CHIME/FRB beam response is a combination of the synthesized beam pattern described in Section~\ref{sec:fftbeamforming} and the primary beam response. The fundamental primary beam response is that of a single feed over a cylindrical reflector, which varies smoothly as an elliptical shape spanning ${\sim}120^{\circ}$ North-South along the local meridian and ${\sim}2.5-1.3^{\circ}$ East-West ($400-800$\,MHz). See Figures~19$-$21 of \cite{chime_overview} for representative plots of primary beam response. However, in practice, the primary beam exhibits more complicated variations resulting from reflections within the telescope and cross-talk between neighboring feeds on the focal line. One of the strongest components of these variations is a $30$\,MHz ripple in the primary beam response as a function of frequency on the order of ${\sim}30{-}50\%$ in amplitude, caused by so-called ``standing wave'' reflections between the focal line and the cylinder \citep{briggs1997,pb08,bna+16}. The form of this ripple is dependent on zenith angle, as shown in Figure~\ref{fig:primary_beam_profiles}.

The shape of the primary beam ripple is also affected by the interferometric complex gain calibration process completed upstream from the FFT beamforming and the CHIME/FRB backend. This gain calibration process uses the transit of a compact steady source of known flux (e.g., Cygnus A, Cassiopeia A, or Taurus A) to correct for signal delays introduced in the analog chain between the feeds and the correlator \citep{chime_overview}. This process essentially normalizes the primary beam attenuation to unity for all frequencies at the meridian transit location of the calibration source. As a result, sources detected at or near the location of the complex gain calibrator will show their true spectrum, while away from the calibrator the detected spectrum will be modulated by the $30$\,MHz ripple. This is demonstrated in Figure~\ref{fig:primary_beam_profiles}, where the spectrum for Cygnus A shows minimal ripples, as it was used as the complex gain calibrator at the time the data for this plot were taken.

\subsubsection{Beam Model} \label{sec:beammodel}
The last two sections have demonstrated the complex nature of the CHIME/FRB beam pattern, characterized by rapidly varying sensitivity across both field of view and bandwidth. Developing an accurate model for such a complex beam is important for separating beam attenuation from intrinsic FRB features to obtain accurate flux and fluence measurements for CHIME/FRB. The CHIME/FRB beam model exists as its own code repository, developed by members of both the CHIME/FRB and CHIME/Cosmology research teams.\footnote{This beam model is available at: \href{https://chime-frb-open-data.github.io/beam-model/}{https://chime-frb-open-data.github.io/beam-model/}}

The clamping patterns from FFT beamforming are digital and deterministic, so modeling them simply involves re-calculating the beamforming process. As a result, an accurate model of the FFT-beamformed beams was available early in the commissioning of the flux calibration pipeline, by July 2018. Modeling the primary beam, however, is much more challenging, as it involves accounting for complex physical reflections and cross-talk within the CHIME cylinders. Therefore, a full sky beam model requires phenomenological and analytical models, informed by the restricted beam measurements via transiting point sources. As a result, for the first year of CHIME/FRB's operation, the primary beam model was just a smooth cosine in the North-South direction and a Gaussian in the East-West direction, with no attempt to characterize the $30$\,MHz ripple. Over the course of the first year of operation, the CHIME/Cosmology collaboration better characterized the $30$\,MHz ripple using a combination of an analytical model and steady source transits \citep{chime_overview}, resulting in a new beam model accurate to within ${\sim}10\%$ in the main lobe of the primary beam \citep{catalog1,chime_sun_measurements}. This data-driven primary beam model was incorporated into the CHIME/FRB codebase in late 2019, after the first catalog of FRBs was detected and the automated fluence pipeline was developed.

\subsection{Localization Limitations} \label{sec:localization_limitations} 

The automated form of localization developed for CHIME/FRB bursts detected with only intensity data is ``header localization,'' which works by fitting the detection S/N from each beam to predictions informed by the frequency-dependent beam model (see Section~3.1 of \citealt{rn1} for a more detailed description). Note that the CHIME/FRB baseband localization system, capable of localizing high-S/N bursts with subarcminute precision, was developed later in CHIME/FRB commissioning \citep{baseband_paper}. In addition, baseband data are only available for $146$ of the $536$ bursts in the first CHIME/FRB catalog \citep{catalog1}.

\begin{figure} 
      \centering
      \includegraphics[width=0.8\columnwidth]{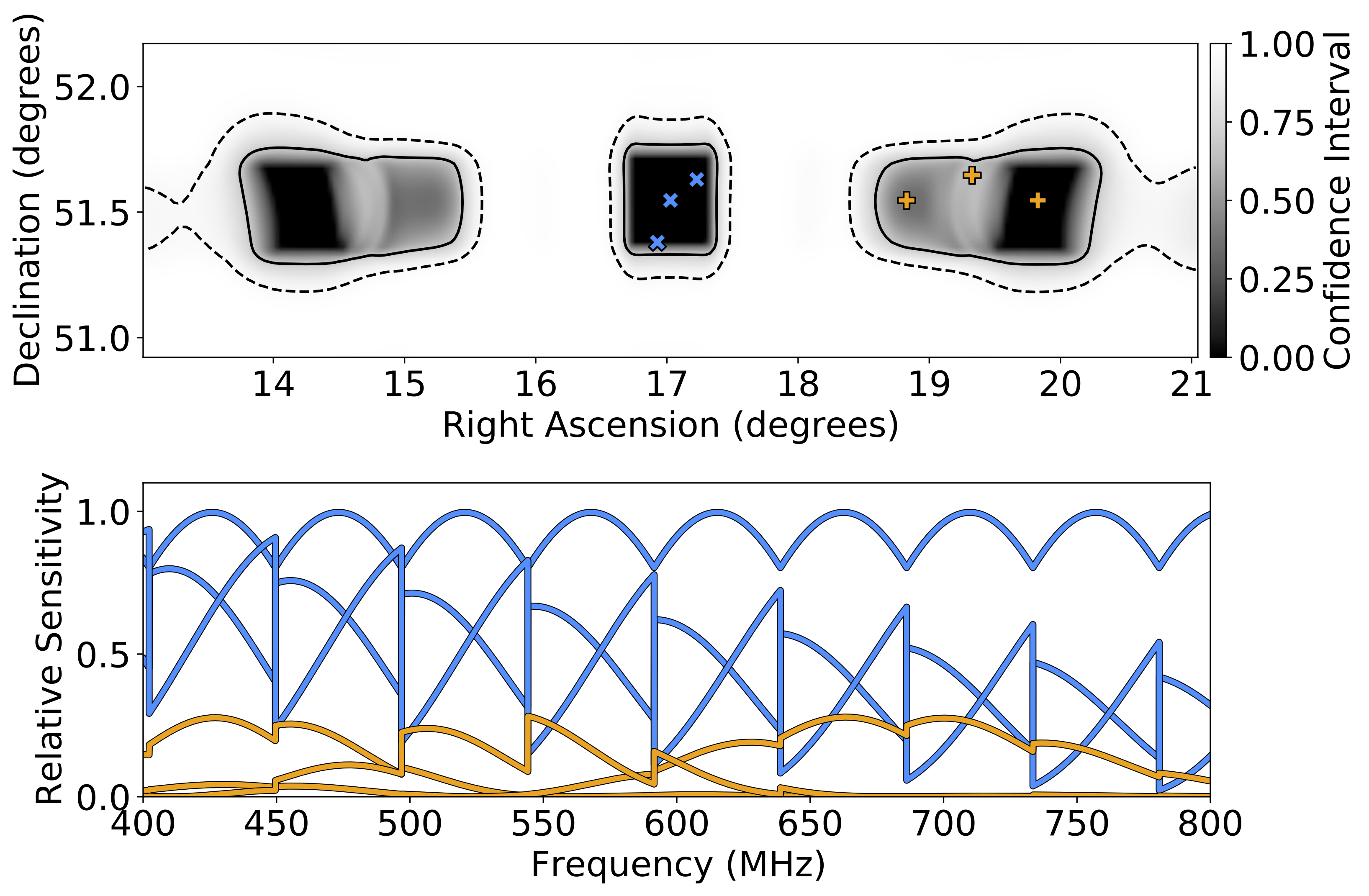}
      \caption{(Top) An example header localization confidence interval for a single beam detection. The $68\%$ and $95\%$ intervals are labelled by the solid and dashed contours, respectively. The blue crosses represent sampling locations in the main lobe of the localization, while the orange plus signs represent sampling locations in the sidelobes. (Bottom) The CHIME/FRB beam response as a function of frequency at the samplings marked in the top panel. Note that these beam responses were generated using a smooth approximation of the primary beam model (see Section~\ref{sec:primarybeam}). Blue lines correspond to the main lobe samplings, orange lines correspond to the sidelobe samplings.}\label{fig:header_loc_sampling_freq}
    \end{figure}

Owing to CHIME/FRB's significant and highly chromatic formed-beam sidelobes, the confidence regions from header localization generally consist of three ``islands'' representing the main lobe and sidelobes of the highest SNR formed beam in which the burst was detected (see the top panel of Figure~\ref{fig:header_loc_sampling_freq}). Multi-beam detections result in better localizations depending on the pattern of detected beams, but generally the degeneracy between the main lobes and sidelobes persists, especially at higher confidence levels. Given this degeneracy, the spectral structure of CHIME/FRB's beam pattern and overall beam response can change significantly over the extent of the header localization region obtained for each burst, making it difficult to reliably correct fluence measurements for beam attenuation. Figure~\ref{fig:header_loc_sampling_freq} shows the beam response at several locations within the $68\%$ confidence localization region for a single-beam detection, demonstrating how the clamping behavior and overall beam response can change rapidly both within the main lobe and within the sidelobes of the localization. 
    
\section{The Automated Calibration Pipeline} \label{sec:automated_calibration_pipeline}
\begin{figure}
  \centering
  \includegraphics[width=1.\columnwidth]{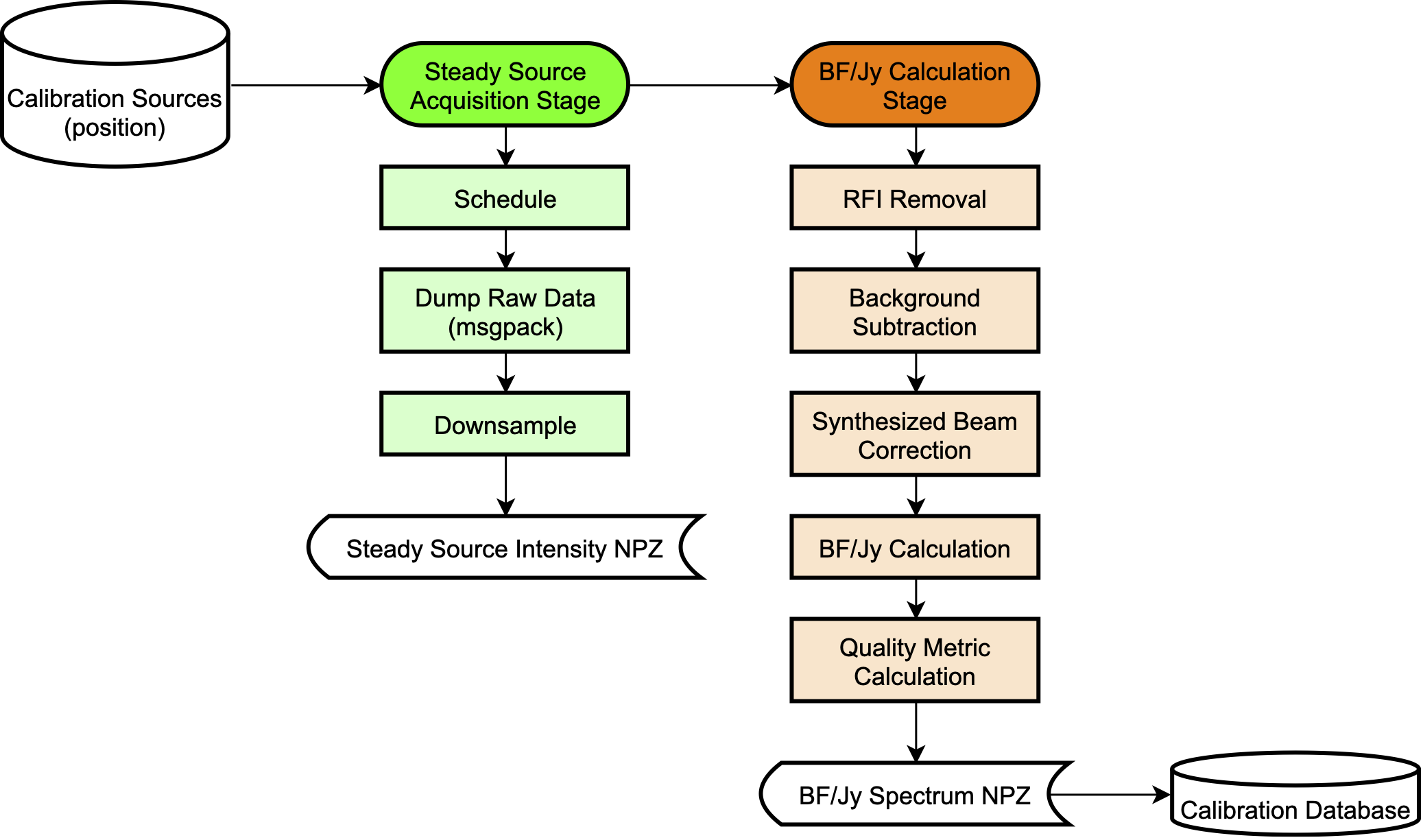}
  \caption{A flowchart of the first half of the CHIME/FRB flux calibration pipeline, which encompasses steady source acquisitions and determination of BF/Jy conversion spectra.}\label{fig:cal_pl_steady_source}
\end{figure}

The primary question motivating the design of the flux calibration pipeline is the following: given the beam modeling and localization limitations during CHIME/FRB's commissioning phase, how do we leverage our existing resources to obtain meaningful constraints on the flux and fluence of each detected FRB? Answering this question prompted two salient design decisions:
\begin{enumerate}
    \item Since we did not have an accurate model of the primary beam when the calibration pipeline was first developed, we characterize and correct for the $30$\,MHz ripple empirically using daily transit observations of steady sources with known spectral properties. By comparing the known flux of a source to the observed total-intensity units output by the beamformer (BF), we solve for the beamformer-to-Jansky (BF/Jy) conversion across the primary beam directly rather than relying on measurements of the system temperature, gain, or SEFD in combination with an approximate beam model. Determining this conversion for each frequency channel creates a ``calibration spectrum'' that encodes the $30$\,MHz ripple in the direction of a given calibration source. This spectrum is then applied to the total-intensity data of FRBs nearby in zenith angle to derive a burst dynamic spectrum in physical units roughly corrected for attenuation from the North-South pattern of the primary beam.
    
    \item Due to the limitations of header localization combined with CHIME/FRB's complex sensitivity pattern, we follow suit from other early FRB surveys and calculate our fluences assuming that each burst was detected at beam boresight. For our purposes we take ``boresight'' to mean along the meridian of the primary beam (at the peak sensitivity of the burst declination arc, in the 1000 beam column). We do not correct our fluence measurements for a burst's unknown location in the synthesized beam pattern. Thus, our fluence measurements are biased low, as bursts off-meridian will experience beam attenuation from both the primary and synthesized beam pattern that we are not correcting for. The measurements produced from the pipeline are therefore most appropriately interpreted as \textit{lower limits}, with an uncertainty on the limiting value. 
\end{enumerate}
In this section we present the implementation details of the current CHIME/FRB flux calibration pipeline. In Section~\ref{sec:pipeline_infrastructure} we give an overview of the technical infrastructure of the pipeline. Sections~\ref{sec:steady_source_acquisition_stage} and \ref{sec:bf_to_jy_stage} describe the first half of the pipeline outlined in Figure~\ref{fig:cal_pl_steady_source}, encompassing steady source observations and determination of the BF/Jy conversion spectra. Sections~\ref{sec:intensity_calibration_stage} and \ref{sec:fluence_calculation_stage} describe the second half of the pipeline outlined in Figure~\ref{fig:cal_pl_flux_cal}, encompassing the intensity calibration and fluence calculation stages. Finally, Section~\ref{sec:comparison_with_injections} presents a test of the flux calibration pipeline, which compares measured fluence values from the pipeline to the values of injected bursts of known fluence.

\subsection{Pipeline Infrastructure} \label{sec:pipeline_infrastructure}

Since CHIME/FRB detects a large volume of bursts, the flux calibration pipeline is automated as much as possible to save human work hours. The BF/Jy calculation, intensity calibration, and flux calculation stages are configured to run in jobs distributed on a separate $10$-node on-site compute cluster. Each of the nodes in this cluster consists of a dual socket Intel\textsuperscript{\textregistered} Xeon\textsuperscript{\textregistered} CPU E5-2630 v4 2.20 GHz with 128 GB of RAM. Using a virtualization platform called Docker,\footnote{\url{https://www.docker.com/}} the flux calibration pipeline code is organized into a self-contained software package called a container, which includes all of the libraries, system tools, and dependencies needed to run the code in any compute environment. This container can be launched onto the cluster to run any stage of the flux calibration pipeline on any detected FRB event. The cluster is managed by a custom-built load balancing service using Docker Swarm.\footnote{\url{https://docs.docker.com/get-started/orchestration/}} Each calibration pipeline job is allocated a single core and $16$\,GB of RAM. This setup allows multiple jobs to be run on the cluster in parallel, which is conducive for analyzing large batches of bursts. 

When an FRB is detected in the CHIME/FRB real-time detection pipeline \citep{chimefrb_overview}, the burst is assigned a unique event ID number and beamformed intensity data containing the burst is saved in an on-site CHIME/FRB archiver with $450$ TB of storage. Once an FRB has been verified through human inspection, several automated post-detection analyses are triggered in order to better characterize the burst properties: 1) the dispersion measure (DM) pipeline determines the S/N- and structure-optimized DMs; 2) the localization pipeline uses real-time detection metadata to calculate the header localization confidence region (as described in Section~\ref{sec:localization_limitations}); and 3) \texttt{fitburst} software is used to model the burst morphology in time and frequency space and determine fundamental burst parameters like the DM, time of arrival (TOA), intrinsic width, and scattering time (for more information about \texttt{fitburst}, see \citealt{abb+19a}, \citealt{abb+20}, or \citealt{fab+20}). The results from each pipeline are stored in a central CHIME/FRB Event Parameters Database, implemented with the \texttt{MongoDB} database management program.

The flux calibration pipeline is triggered after the \texttt{fitburst} pipeline. Flux calibration pipeline data products --- such as the downsampled calibration source intensity data, BF/Jy conversion spectra, and calibrated FRB dynamic spectra --- are stored in the CHIME/FRB archiver. Output fluence measurements are also stored in the Event Parameters Database, while metadata for calibration data products are stored in a separate Calibration Database. These databases can be queried from anywhere through authenticated HTTP requests using a RESTful Python API. Querying of the Calibration Database is designed to be flexible, allowing the user to obtain the available calibration spectra on a given date, in a date range, from a given calibration source, or nearest to a spatial location on-sky. This functionality is used extensively in the intensity calibration and fluence calculation stages of the pipeline.

\subsection{Steady Source Acquisition Stage} \label{sec:steady_source_acquisition_stage}
The first stage of the fluence pipeline (the ``Steady Source Acquisition Stage'' in Figure~\ref{fig:cal_pl_steady_source}) deals with scheduling, processing, and storing daily total-intensity transit observations of steady calibration sources. 

\subsubsection{Steady Source Selection}

Calibration sources were selected from two different catalogs, \cite{perley:2016} and \cite{vollmer:2010}, which provide measurements of the source flux across the CHIME band. \cite{perley:2016} present fluxes from the Karl G.~Jansky Very Large Array (VLA) measured in 2014 and 2016 at frequencies ranging from $220$\,MHz to $48.1$\,GHz. Combining these measurements with $73.8$\,MHz legacy observations from 1998, \cite{perley:2016} model the frequency dependence of the flux density of each calibrator by fitting a polynomial to the logarithm of the observed flux densities. The resulting uncertainty on the flux of these sources in the CHIME band is ${\sim}2{-}4\%$. \cite{vollmer:2010} present a compilation of flux densities at frequencies ranging from $159$\,MHz to $8.4$\,GHz, derived from cross-identifying sources in several radio catalogs. Since these measurements come from a wide range of telescopes observing at different epochs, the uncertainties on the flux densities in the CHIME band are larger than those from \cite{perley:2016}, on the order of ${\sim}5{-}15\%$.

Sources were selected from these catalogs according to the following three criteria: (1) the source is within the CHIME field of view (above $-11^{\circ}$ declination), (2) the flux density of the source at $600$\,MHz is greater than $10$\,Jy (below which confusion noise becomes significant), and (3) the catalogs provide flux density measurements for at least three different radio frequencies spanning the CHIME band. A total of $35$ sources matched these criteria ($14$ from \citealt{perley:2016} and $21$ from \citealt{vollmer:2010}). 

Note that there are a few sources in our calibrator sample, such as TauA, CasA, and 3C138, that are known to be variable on the order of a few percent per year. For example, TauA's flux density is declining by approximately $0.25$\% every year and 3C138 is declining on the order of ${\sim}1{-}3\%$ per year \citep{baars:1977,perley:2016}. This decline is much smaller than the uncertainties on the flux expected from beam attenuation and the time-variable system sensitivity, and so we keep these sources in our sample to achieve more coverage of the primary beam (see the discussion of uncertainty calculation in Section~\ref{sec:fluence_calculation_stage}).

\subsubsection{Spectrum Interpolation to the CHIME Band}
We use the same methodology as \cite{perley:2016} to interpolate the spectrum of each source and determine the flux in the CHIME band (see Section~4 of \citealt{perley:2016}). The frequency dependence of the flux density of each calibrator is modeled with the following function:
\begin{align}
\log{\left[S(\nu)\right]} & = \sum_{i=0}^{N-1} c_{i} \left[ \log{\left(\frac{\nu}{\nu_{0}} \right)} \right]^{i}
\end{align}
where $\nu$ is the frequency, $S(\nu)$ is the flux density, $\nu_{0}$ is a fixed pivot frequency, $c_{i}$ are the model parameters, and $N$ is the number of parameters. This model is fit to measurements of the flux density by applying weighted least squares to the log-transformed data. The best-fit model is then used to predict the flux of the calibrator in the $400-800$\,MHz CHIME band.

The \cite{perley:2016} catalog does not provide the underlying measurements of the flux density made by the VLA, so for calibrators that originate from \cite{perley:2016} we use the best-fit coefficients that are provided in that work. For these sources $\nu_{0}$ is set to $1$~GHz and $N$ ranges from $3$ to $6$. The uncertainty on the predicted flux density is obtained by propagating the uncertainty on the coefficients, ignoring any covariance between coefficients since this is not provided.

The \cite{vollmer:2010} catalog provides the underlying measurements of the flux density, so we perform the fit ourselves for these sources. The pivot frequency $\nu_{0}$ is set to the center of the CHIME band (600 MHz). A simple power-law ($N=2$) yields an acceptable fit for all sources. The uncertainty on the predicted flux density is obtained by propagating the uncertainty on the individual measurements.

\subsubsection{Observation Scheduling}
The steady source acquisition stage runs continuously in the background on an on-site compute node. Taking the calibrator locations and observation duration as input, this stage uses \texttt{pyephem}\footnote{\url{https://rhodesmill.org/pyephem/}} in combination with the FFT-formed beam model to predict when a given source will transit and which beam in the 1000 column it will cross. When a source is close to transiting, the script creates a folder on the CHIME/FRB archiver, triggers an intensity dump in the transit beam, waits the observation duration, and then stops the dump. Although it only takes a source ${\sim}5{-}14$ minutes to pass through the main lobe of the primary beam, each calibrator is observed for one hour centered around transit in order to obtain a measurement of the adjacent sky background. Note that, for circumpolar calibrators, we only take observations of primary (upper) transits.

The raw intensity data from these observations are output in \texttt{msgpack}\footnote{\url{https://msgpack.org/index.html}} format with $16{,}384$ frequency channels ($24$\,kHz resolution) and $0.983$\,ms time resolution, where each msgpack stores one second of data. A full hour-long observation at this resolution takes up $60$\,GB of memory on the archiver. To save space, the raw data from each observation are unpacked, downsampled in time to $1$-s resolution (taking the median over $1{,}024$ $1$-ms samples for each new $1$-s sample), and organized into a 2D array. The final result is a $200$\,MB \texttt{numpy} NPZ file\footnote{\url{https://imageio.readthedocs.io/en/stable/format_npz.html}} representing the dynamic spectrum of the observation. After the NPZ has been saved, the raw data are deleted.

\subsection{Beamformer Unit to Jansky Calculation Stage} \label{sec:bf_to_jy_stage}
The second stage of the fluence pipeline (the ``BF/Jy Calculation Stage'' in Figure~\ref{fig:cal_pl_steady_source}) deals with extracting a BF/Jy conversion spectrum from each steady source observation, as well as additional metrics.

The first step in this process involves removing radio frequency interference (RFI) from the steady source spectra. CHIME observations are affected by RFI mainly from the LTE band (${\sim}700-800$\,MHz) and transiting airplanes. The data are run through three RFI filters in the flux calibration pipeline. One is a static mask that removes channels that are consistently contaminated. The second stage detects any additional bad frequency channels using a sliding window algorithm that calculates median values of the median absolute deviation and kurtosis over a given frequency channel and time window, and then flags outlier channels based on empirical thresholds. The final RFI removal stage searches for bad time bins by taking the gradient of the observation time series and flagging significant spikes. Altogether this RFI method is rather aggressive. It produces clean data but removes a significant fraction of CHIME's bandwidth, typically leaving ${\sim}260$\,MHz of usable bandwidth.

\begin{figure}[p!]
  \centering
  \includegraphics[width=1.\columnwidth]{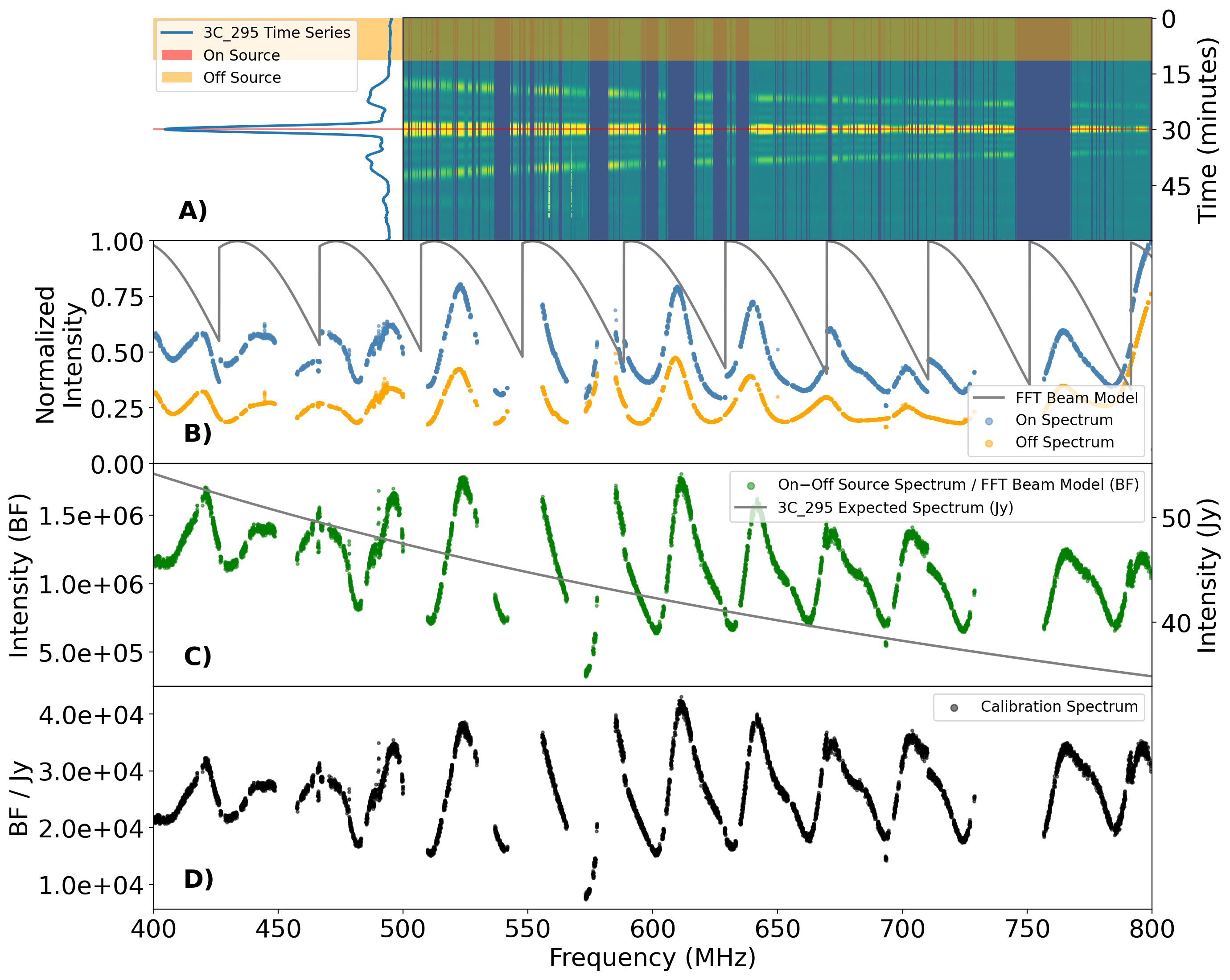}
  \caption{A series of plots showing different stages in the process of calculating a BF/Jy spectrum. The data in this figure come from a March 22nd, 2019 observation of Seyfert 2 galaxy 3C295 transiting through beam number 1134. Panel A shows the dynamic spectrum of the observation (Right), along with the frequency-summed time series (Left). For the dynamic spectrum, the x-axis corresponds to frequency ranging from 400 to 800\,MHz, left to right, while the y-axis corresponds to time. The yellow region denotes the time bins chosen to represent the off-source spectrum, while the red region denotes the time bins chosen to represent the on-source spectrum. Panel B shows the resulting on-source spectrum (blue dots), off-source spectrum (orange dots), and sensitivity of the formed beam at the location of the source during transit (grey solid line). Panel C shows the on$-$off source spectrum divided by the formed beam sensitivities in beamformer units (green dots) as well as the known spectrum of the source in Jy (grey solid line). The left-hand axis shows the source spectrum intensity in beamformer units, while the right-hand axis shows the known source spectrum intensity in Jy. Panel D shows the final resulting BF/Jy spectrum.}\label{fig:bf_to_jy_diagnostic}
\end{figure}

After RFI removal, the BF/Jy spectrum is calculated. This multistep process, outlined in Figure~\ref{fig:bf_to_jy_diagnostic}, can be described by the single equation:
\begin{equation}
    C_{\nu, \text{BF/Jy}} = \frac{\left(S_{\nu, \text{cal}, \text{on}} - S_{\nu, \text{cal}, \text{off}}\right) [\text{BF}]}{\left(B_{\nu, \text{FFT}}(\theta, \phi) \times S_{\nu, \text{cal}, \text{known}}\right) [\text{Jy}]}
\end{equation}
where $C_{\nu, \text{BF/Jy}}$ is the resulting BF/Jy calibration spectrum, $S_{\nu, \text{cal}, \text{on}}$ is the intensity spectrum detected in beamformer units at the peak of the calibrator transit, $S_{\nu, \text{cal}, \text{off}}$ is the background intensity spectrum detected in beamformer units before or after the calibrator has transited, $B_{\nu, \text{FFT}}(\theta, \phi)$ is the synthesized beam sensitivity at the location that the calibrator reaches during peak transit (derived from the formed beam model), and $S_{\nu, \text{cal}, \text{known}}$ is the known flux spectrum of the source modeled as described in Section~\ref{sec:steady_source_acquisition_stage}. The resulting BF/Jy spectrum is saved in an NPZ file on the archiver, and metadata about the spectrum are sent to the Calibration Database. These metadata include descriptive information (e.g., the name of the calibrator, the date of the observation, the beam that the source transits through, and the file path of the BF/Jy spectrum NPZ on the archiver). 

\begin{figure}[!ht]
  \centering
  \includegraphics[width=.8\columnwidth]{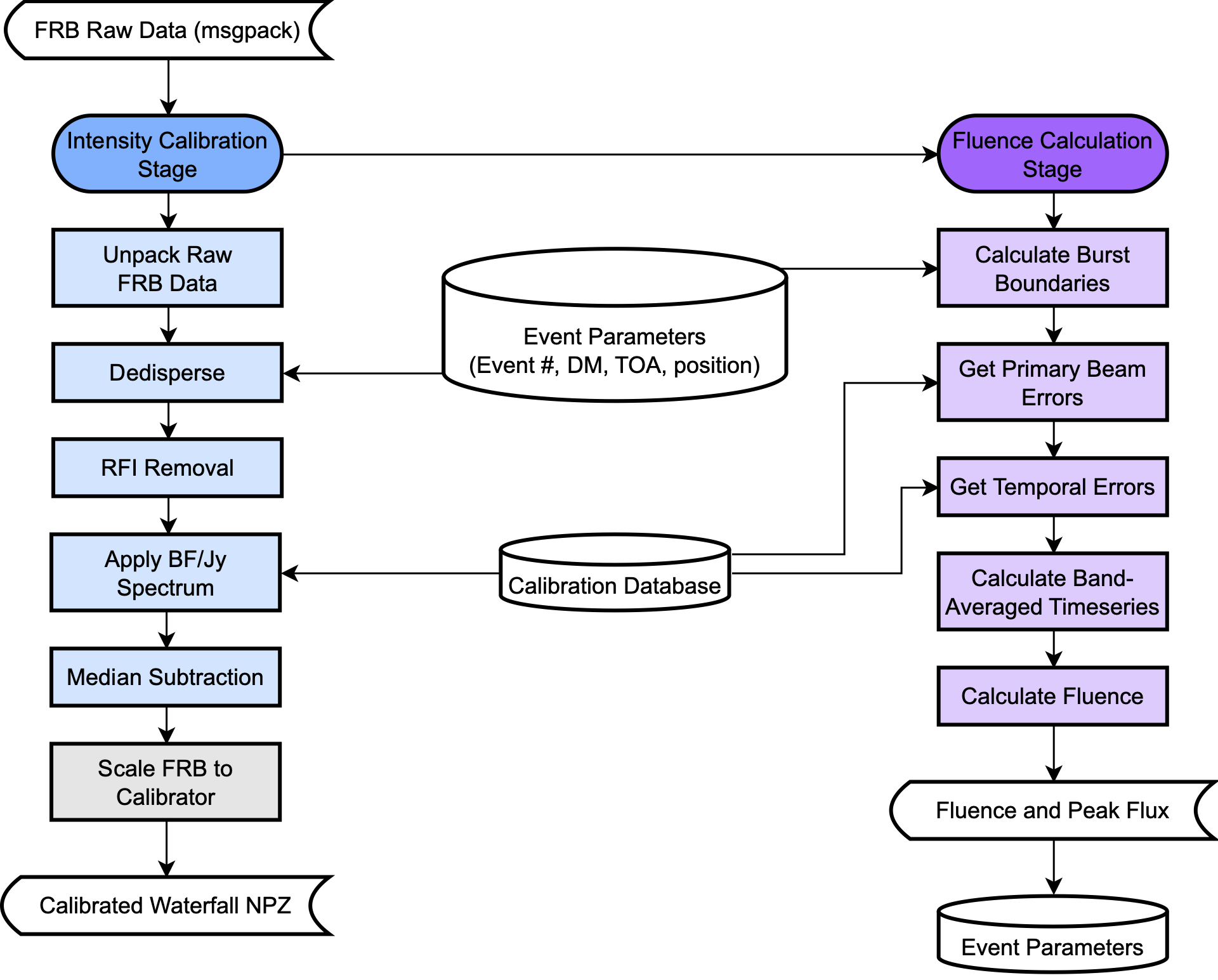}
  \caption{A flowchart of the second half of the CHIME/FRB flux calibration pipeline, which encompasses the intensity calibration and flux calibration calculations. Optional steps, that are not always executed, are greyed out. }\label{fig:cal_pl_flux_cal}
\end{figure} 

\subsection{Intensity Calibration Stage} \label{sec:intensity_calibration_stage}
The second half of the flux calibration pipeline, outlined in Figure~\ref{fig:cal_pl_flux_cal}, switches focus from processing calibrator observation data to processing FRB data. The ``Intensity Calibration Stage,'' in particular, deals with applying a BF/Jy spectrum to FRB intensity data to obtain a calibrated dynamic spectrum (``waterfall'').

Similarly to the steady source transit observations, raw FRB intensity data are saved on the archiver in msgpack format with $16{,}384$ frequency channels ($24$\,kHz resolution) and $0.983$\,ms time resolution. For each FRB that enters the flux calibration pipeline, msgpack data are unpacked, organized into a 2D dynamic spectrum, have RFI excision applied, and de-dispersed to the best-fit DM derived from \texttt{fitburst}. 

After preprocessing, the FRB data are then calibrated. Each burst is paired with the calibration spectrum of the nearest steady source transit, closest first in zenith angle, then in time. We assume North-South beam symmetry, so that sources on both sides of zenith can be used to calibrate a given event. By dividing each frequency channel in the intensity data (in beamformer units) by its corresponding BF/Jy conversion, we derive a dynamic spectrum in physical units (Jy) roughly corrected for North-South beam variations. As a final step, each frequency channel is subtracted by its median value. The resulting dynamic spectrum is saved in an NPZ on the archiver.

\begin{figure}[p!]
  \centering
  \includegraphics[width=.8\columnwidth]{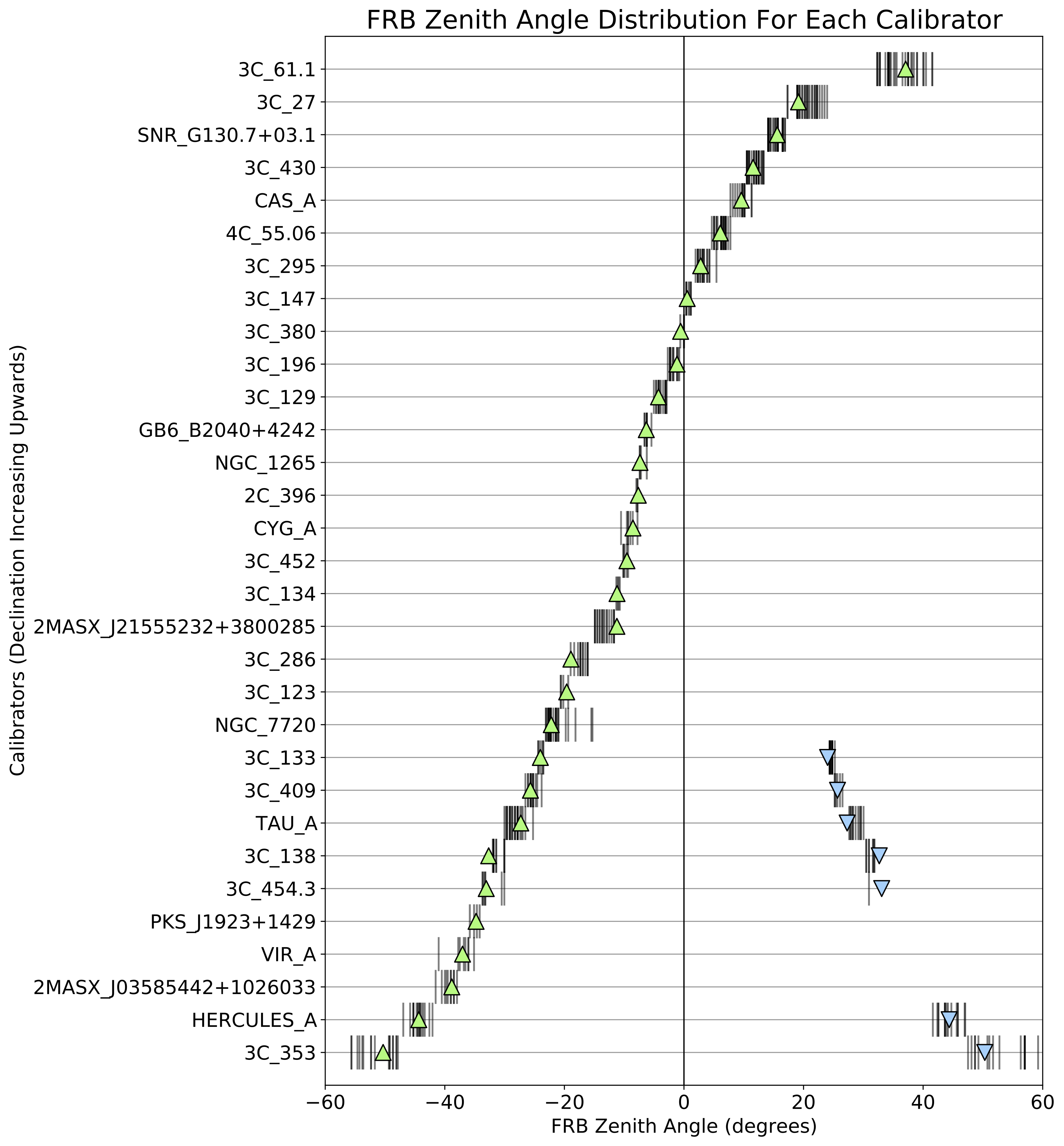}
  \caption{The association of FRBs with calibrators as a function of zenith angle for the first CHIME/FRB catalog \citep{catalog1}. Each calibrator is labelled on the y-axis and represented by a horizontal line, with the y-axis sorted by increasing calibrator declination. The position of each calibrator is marked with an upward pointing green triangle. The symmetric position on the opposite side of zenith is marked with a downward pointing blue triangle. An FRB paired with a given calibrator is marked on the horizontal line of the calibrator by a small vertical line at the zenith angle of the FRB. }\label{fig:calibrators_and_frb_zenith_angle}
\end{figure}

Figure~\ref{fig:calibrators_and_frb_zenith_angle} shows a spatial representation of the calibrator-FRB pairings for the bursts in the first CHIME/FRB catalog \citep{catalog1}. $98\%$ of FRBs are associated with a calibrator within $5$ degrees in zenith angle (either on the same side of zenith or at North-South symmetric locations on opposite sides of zenith). Most of the remaining $2\%$ of FRBs were detected at the edges of CHIME's North-South field of view (very high or low zenith angles), where 3C353 is the only available calibrator. Especially at high northern zenith angles, there is a paucity of calibrators since we do not currently dump observations for the lower transits of circumpolar sources. Note that these calibrator-FRB pairings are a function of spatial and temporal proximity as well as data quality. As a result, not all FRBs are associated with their nearest calibrator, as the data for that calibrator could be disrupted by solar transits for a couple of weeks at a time. This is the case for two FRBs detected $6{-}7$ degrees away from NGC7720, that are still paired with the source for calibration.

As discussed in the introduction of Section~\ref{sec:automated_calibration_pipeline}, by default, we do not correct FRBs for attenuation due to their uncertain locations within the synthesized beam. However, after a more accurate model of the primary beam was developed in late 2019, we added an option to the pipeline that produces a per-frequency scaling between the location of the calibrator and the location of the FRB using the beam model. This is useful functionality for when a burst has a precise localization determined using CHIME/FRB baseband or another long-baseline telescope. In these instances, we can apply this scaling to the dynamic spectrum to obtain accurate fluence and flux measurements corrected for synthesized beam attenuation, rather than lower bounds. This method was used to analyze the fluence distribution of the first repeating FRB with periodic activity, FRB 20180916B, which had a precise localization from the European VLBI Network \citep{aab+20}. 

\subsection{Fluence Calculation Stage} \label{sec:fluence_calculation_stage}
The final stage of the calibration pipeline deals with the calculation of FRB fluence and flux values from the calibrated dynamic spectra (the ``Fluence Calculation Stage'' in Figure~\ref{fig:cal_pl_flux_cal}).

The first step in the fluence calculation involves determining the boundaries of the burst extent in the dynamic spectrum. The fluence and flux values calculated from the pipeline are averaged over the entire $400{-}800$\,MHz CHIME band. For narrow-band bursts, this averages noise into our fluence and flux values. However, we choose to quote values from the same frequency range for consistency. In terms of determining the burst boundaries, this means that we only need to localize the burst along the temporal axis of the dynamic spectrum. This is accomplished using results from the \texttt{fitburst} routine, which outputs fundamental burst parameters like the arrival time $t_{\text{arr}}$, intrinsic width (the Gaussian $\sigma$), and the scattering time $\tau$ at $400$\,MHz. The start and end times encompassing the burst are defined by the $3\sigma$ Gaussian width, with an optional extra term added to the end time to account for a scattering tail, if present:
\begin{align}
    t_{\text{start}} &= t_{\text{arr}} - 3\sigma \\
    t_{\text{end}} &= t_{\text{arr}} + 3\sigma + 5\tau \cdot (p < 0.001)
\end{align}
where $p$ is the p-value from an F-test comparison between scattered and un-scattered models fit to the burst. A p-value significance of $0.1\%$ was used to declare the presence of significant scattering. In this case, we add five times the scattering time to the time denoting the end of the burst, corresponding to the time where the scattered emission would have decreased by a factor of $e^{-5} \approx 0.008$ (less than $1\%$). Note that these encompassing start and end times can be modified following human inspection.

\begin{figure}
  \centering
  \includegraphics[width=.6\columnwidth]{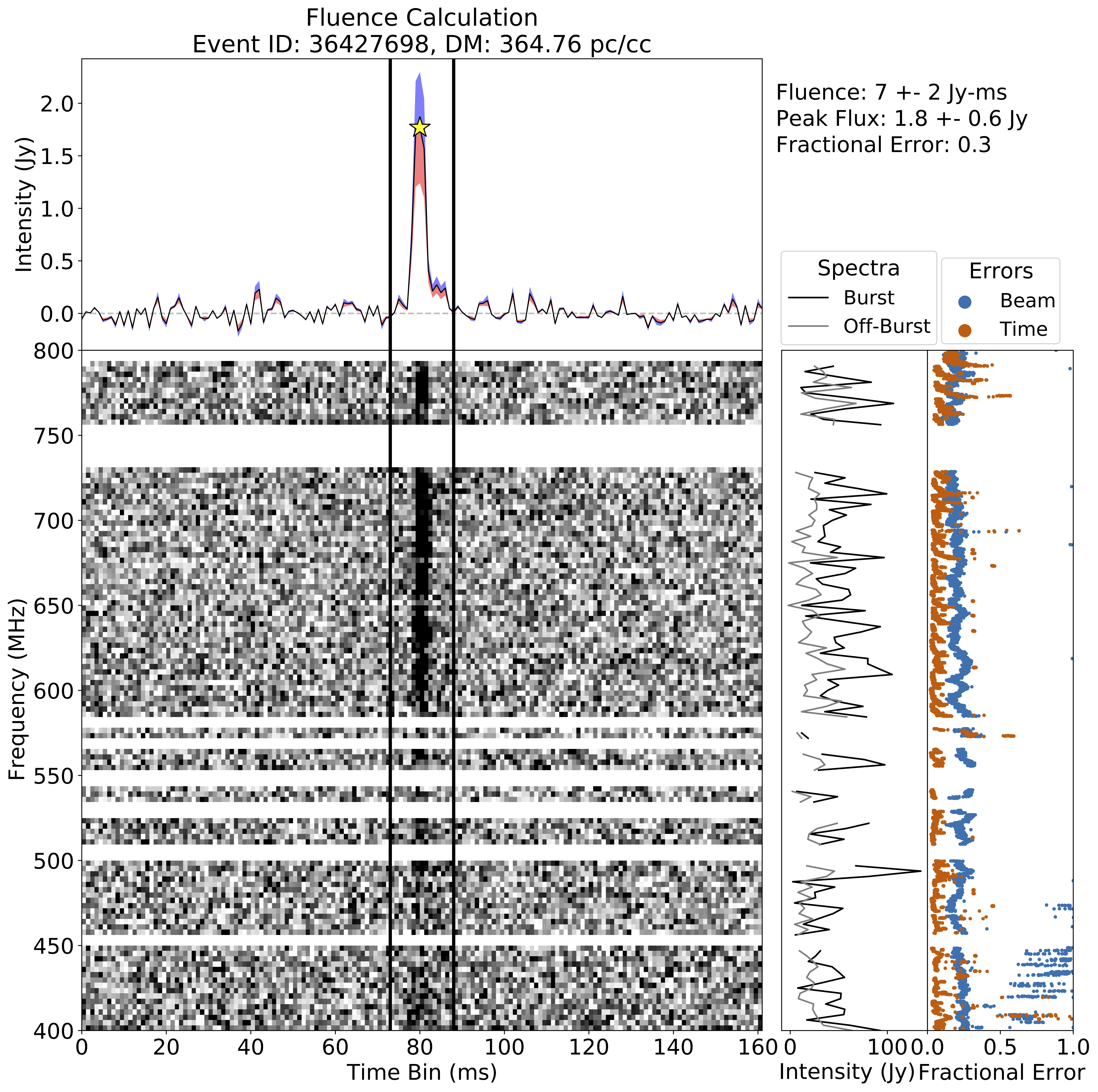}
  \caption{A diagnostic plot for the ``Fluence Calculation Stage'' of the flux calibration pipeline. The central panel is the dynamic spectrum of the burst, where the color scale shows the intensity (Jy) in each frequency-time bin. Note that this dynamic spectrum is subbanded to $128$ frequency channels for plotting clarity, but this is not included as a step in the fluence calculation. The white horizontal sections of the dynamic spectrum indicate subbanded channels that have been completely masked due to RFI. The top panel shows the band-averaged time series of the burst (black line) as well as the upper bound (blue shading) and lower bound (red shading) uncertainty for each time bin. The yellow star indicates the bin from where the peak flux is taken. The vertical lines indicate the extent of the burst integrated over to obtain the fluence. The first panel to the right of the dynamic spectrum shows the subbanded average spectrum of both the burst (black) and the off-burst background (grey). The far right panel shows the fractional error in each frequency bin determined due to variations in time (orange) and across the primary beam (blue).}\label{fig:fluence_diagnostic}
\end{figure}

Next, the band-averaged time series is derived by averaging the de-dispersed calibrated dynamic spectrum over the bandwidth remaining after RFI removal, and subtracting the resulting time series by the median of the off-pulse. The fluence is then calculated by integrating over the burst extent in the band-averaged time series, while the peak flux is the maximum value within the burst extent (at $0.983$\,ms resolution). Figure~\ref{fig:fluence_diagnostic} shows an example band-averaged time series, with the burst extent and peak flux bin labeled.

The uncertainties on the fluence and flux values are estimated using steady source transit data. There are two main sources of error that are incorporated into the pipeline uncertainties: 1) the error due to differences in the primary beam response between the calibrator and the assumed FRB location along the meridian, or the ``primary beam error,'' and 2) the error due to the temporal separation between the FRB and the calibrator transit, or the ``time error'' (this error encompasses temporal variations in the system sensitivity, as well as calibrator source variability). Each of these errors is first calculated as a relative or fractional uncertainty in each frequency channel (see the rightmost panel of Figure~\ref{fig:fluence_diagnostic}).

\begin{figure}[ht!]
  \centering
  \includegraphics[width=0.48\linewidth]{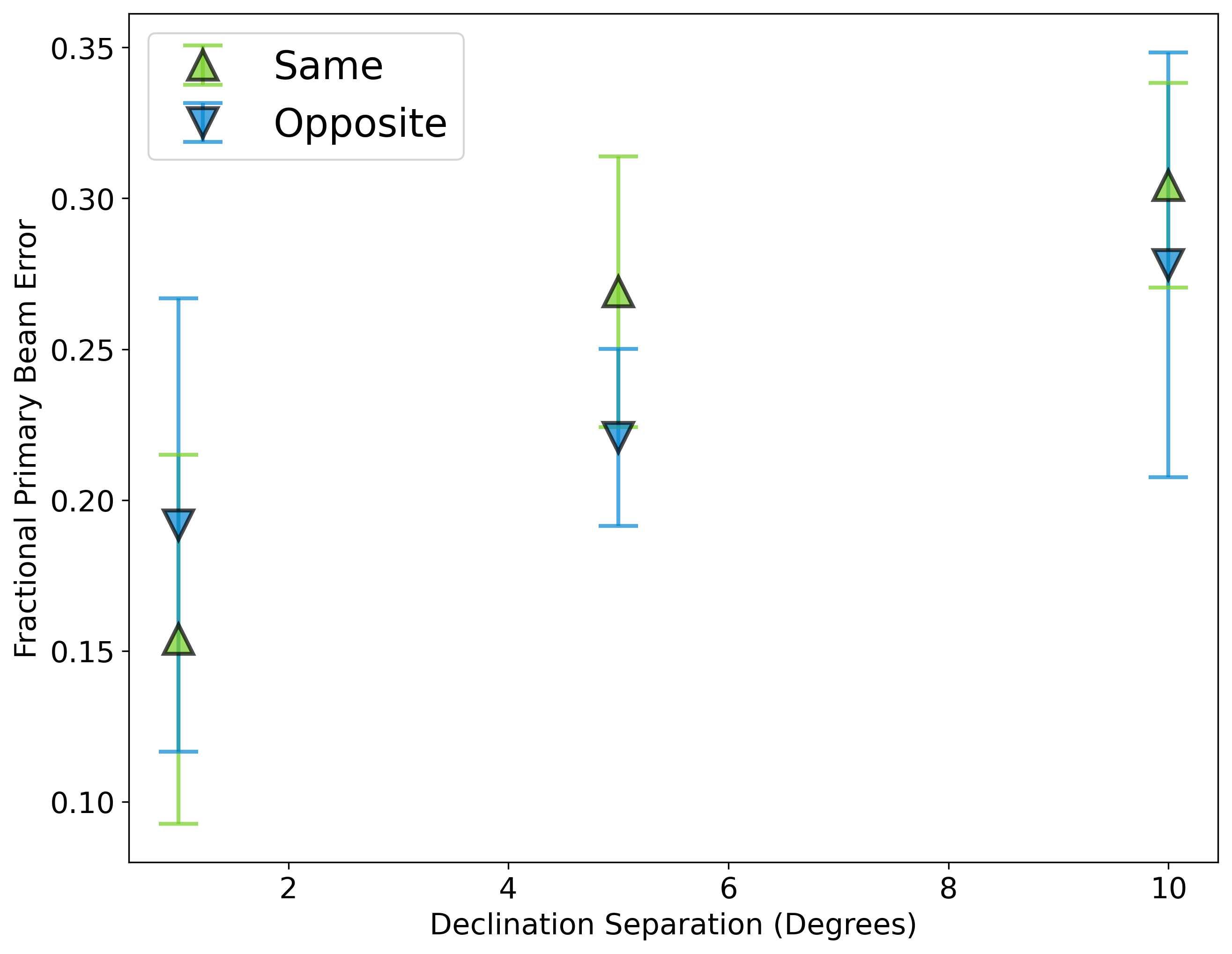} %
  \includegraphics[width=0.48\linewidth]{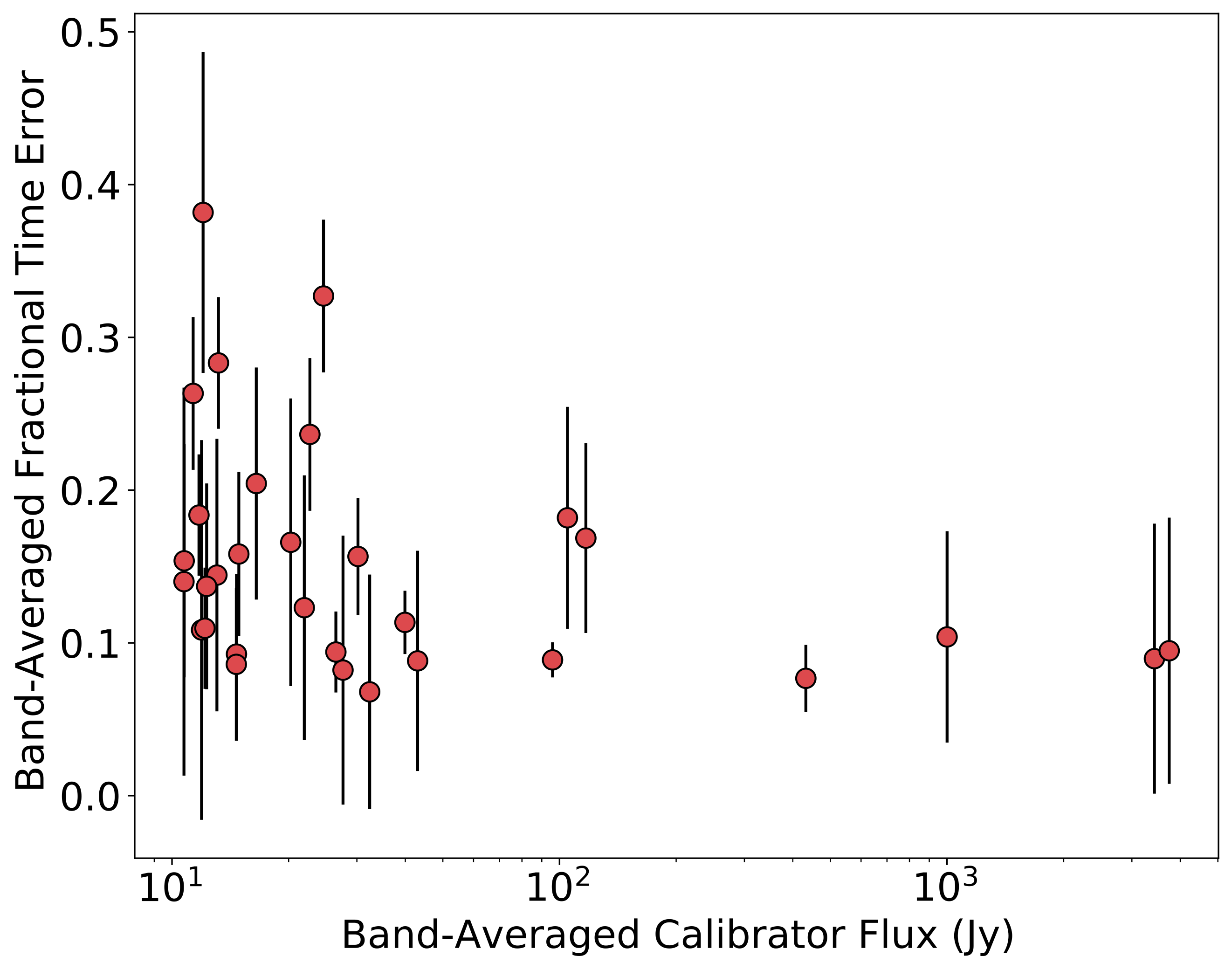}
  \caption{The fractional errors averaged over both bandwidth and bursts in the first CHIME/FRB catalog \citep{catalog1}. (Left) The average fractional error due to the primary beam, split by separation between the FRB and calibrator (points from left to right correspond to: ${<}1^{\circ}$, $1{-}5^{\circ}$, $5{-}10^{\circ}$). Upward pointing green triangles indicate FRB/calibrator pairs on the same side of zenith from each other, downward pointing blue triangles indicate pairs on the opposite side of zenith from each other. Error bars show the standard deviation of the averaged values, with caps that match the color of their associated point. (Right) The average fractional error due to temporal variations, as a function of calibrator flux. Again, error bars show the standard deviation of the averaged values.}\label{fig:errors}
\end{figure}

The primary beam error is estimated by using steady source observations from a single day to calibrate each other and measuring the average fractional error as a function of frequency compared to known flux values (the fractional error is given by: $(S_{\nu, \text{meas}} - S_{\nu, \text{exp}}) / S_{\nu, \text{exp}}$ where $S_{\nu, \text{exp}}$ is the expected spectrum and $S_{\nu, \text{meas}}$ is the measured spectrum). The calibrator pairs used to estimate this error are selected to match the spatial separation in zenith angle between the FRB and its paired calibrator. For example, if an FRB is less than $1^{\circ}$ from its calibrator (in zenith angle), then the primary beam errors are estimated using steady source pairs that are within $1^{\circ}$ of each other. If an FRB is $1{-}5^{\circ}$ from its calibrator, then steady source pairs are selected that are within $5^{\circ}$ of each other. If an FRB is $5{-}10^{\circ}$ from its calibrator, then steady source pairs are selected that are within $10^{\circ}$ of each other. If an FRB/calibrator pair are from opposite sides of zenith, then steady source pairs are selected from a similar distance on opposite sides of zenith. This primary beam error is typically on the order of $20{-}30\%$ (band-averaged), depending on the distance between the FRB and the calibrator. The left panel of Figure~\ref{fig:errors} shows the average fractional primary beam errors derived for different spatial separations.

For a given FRB, the time error is determined by measuring the rms variation in the BF/Jy spectra of the paired calibrator over a period of roughly two weeks surrounding the burst arrival (the fractional error is given by: $\Delta C_{\nu, \text{BF/Jy}} / \overline{C}_{\nu, \text{BF/Jy}}$ where $\Delta C_{\nu, \text{BF/Jy}}$ is the rms of the BF/Jy spectra and $\overline{C}_{\nu, \text{BF/Jy}}$ is the average). The time error is typically on the order of $10{-}20\%$, depending on the calibrator used (see the right panel of Figure~\ref{fig:errors}). 

The time and primary beam fractional errors are combined together to obtain an upper limit on the fractional uncertainty in each frequency channel. Since these errors are systematic and not necessarily Gaussian-distributed, we choose to be conservative and sum them directly rather than add them in quadrature. Multiplying the dynamic spectrum by the combined fractional error along the frequency axis yields an uncertainty on the intensity in each frequency/time bin. We then band-average this uncertainty over the dynamic spectrum, and add and subtract the resulting time series to the previously determined band-averaged time series for the burst. This results in upper bound and lower bound burst profiles, which are indicated by the blue and red shading in the top panel of Figure~\ref{fig:fluence_diagnostic}. The error on the fluence is given by the average area between the upper and lower profiles, and the error on the peak flux is determined from the bounds of the maximum bin. Both of these errors are summed with the rms fluence and flux from the off-pulse region in the band-averaged time series to form the final error.

We note that the estimated errors do not encapsulate the bias due to our assumption that each burst is detected along the meridian of the primary beam, which causes our fluence measurements to be biased low. As previously mentioned, the measurements produced from the pipeline are most appropriately interpreted as \textit{lower limits}, with an uncertainty on the limiting value.

\subsection{Comparison with Injected Bursts} \label{sec:comparison_with_injections}

Early in the commissioning of the CHIME/FRB experiment, the upstream $N^2$ complex gain calibration process was still in development. Over time, the procedure matured from phase-only calibration (during the pre-commissioning period until September 4th, 2018) to phase and amplitude calibration normalized to the daily transit of CygA (as of early 2019). Applying these phase and amplitude gains to the raw upstream baseband data essentially flux calibrates the data up to a static primary beam model. Then, the process of CHIME/FRB beamforming and upchannelization introduces a series of additional scalings, resulting in the observed ``beamformer units'' of the CHIME/FRB intensity data. The conversion between physical flux units and BF units is given by:
\begin{equation} \label{sec:static_calibration_factor}
    1\,\text{Jy} = \frac{\left(1{,}024 f_{\text{good}}\right)^2 \cdot 128}{4^2 \cdot 0.806745 \cdot 400}\,\text{BF} \approx 26{,}000 f_{\text{good}}^2\,\text{BF}
\end{equation}
where $f_{\text{good}}$ is the fraction of good feed inputs. This number is the only time-variable scaling factor introduced in the beamforming process (it generally varies between $70\%$ and $95\%$, depending on environmental conditions around the telescope). The origin of all of the other factors in this equation are enumerated in detail in \cite{injections_paper}.

On April 20th, 2020, the CHIME/FRB beamforming code was updated to account for the $f_{\text{good}}^2$ scaling, so that CHIME/FRB intensity data taken after this date are calibrated in real-time up to the beam model and the static conversion factor given in Equation~\ref{sec:static_calibration_factor}. Comparing CygA transit data calibrated using this conversion factor to the expected spectrum attenuated by the beam model shows that the calibrated flux is accurate to within ${\sim}5\%$.

This new calibration method provided an avenue for testing the flux calibration pipeline using functionality from the CHIME/FRB injections system \citep{injections_paper}. The CHIME/FRB injections system provides infrastructure for generating synthetic FRBs with user-determined properties, injecting them into the real-time intensity data stream to be searched with the CHIME/FRB backend, and tracking the resulting properties measured from the real-time pipeline. Thus, to test the flux calibration pipeline, we simulate a series of synthetic FRB intensity datasets, run them through the flux calibration pipeline, and compare the resulting measured fluences to the nominal simulated values. 

Synthetic FRBs of given fluences and morphologies (e.g., temporal width, bandwidth, scattering, and spectral index) are generated by \texttt{simpulse}\footnote{\href{https://github.com/kmsmith137/simpulse}{https://github.com/kmsmith137/simpulse}}, which produces dynamic spectra of the intrinsic FRBs in Jy units. Using the scaling factor given in Equation~\ref{sec:static_calibration_factor}, the dynamic spectra are converted from Jy to BF units. Then the FRBs are scaled by the beam attenuation from the CHIME/FRB beam model (including the accurate, data-driven primary beam) given a simulated sky location. Finally, each pulse is injected into an empty intensity dataset representing CHIME/FRB background noise.

\begin{figure}[!htb]
  \centering
  \includegraphics[width=.5\columnwidth]{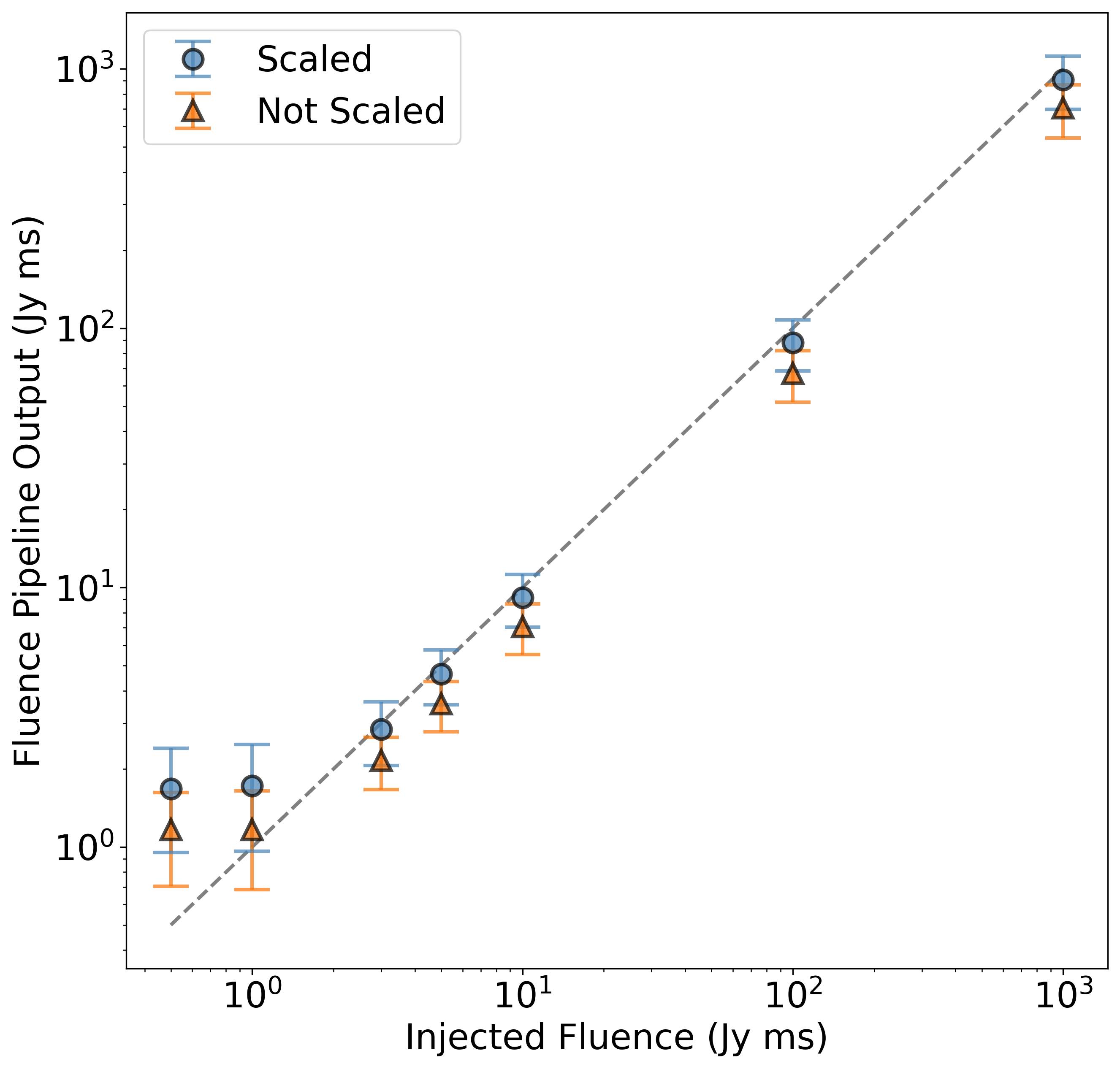}
  \caption{The nominal simulated burst fluence (before beam attenuation) versus the burst fluence derived from the flux calibration pipeline, for a series of bursts injected at the center of formed beam $1070$. The orange triangles indicate pipeline fluences derived without beam model scaling, and the blue circles indicate pipeline fluences derived by scaling between the location of the calibrator and the location of the FRB. The error bars represent the uncertainties derived from the calibration pipeline. The black solid line denotes 1:1. Note that a similar plot was presented by \citep{injections_paper}. We reproduce this plot here, for clarity. }\label{fig:inj_fluence_comparison}
\end{figure}

We simulated a series of simple bursts with no scattering, intrinsic widths of $1$\,ms, flat spectra, and fluences of $0.5$, $1$, $3$, $5$, $10$, $100$, and $1{,}000$\,Jy\,ms. The bursts are injected at the center of formed beam $1070$. Each burst is run through the calibration pipeline twice: first without beam model scaling and then scaling between the location of the calibrator (in this case NGC 7720, which transits through beam $1071$) and the location of the FRB, as described in Section~\ref{sec:intensity_calibration_stage}. The results are displayed in Figure~\ref{fig:inj_fluence_comparison}.

The injected fluence and flux pipeline fluence are in agreement within the errors output by the flux calibration pipeline. As the bursts are injected at the center of a formed beam in the 1000 beam column, these injections represent a ``best case scenario'' for the fluence pipeline, which explains how even the lower-bound fluences are close to the injected values. Note that, at the lowest fluence values of $0.5$ and $1$ Jy, the fluence pipeline reports the fluence of a noise peak instead of the faint synthetic pulse.

\section{Conclusions}

The current flux calibration pipeline described in this paper was developed rapidly to provide flux and fluence measurements for early CHIME/FRB bursts detected with an incompletely characterized system, extending from the first detection on July 25th, 2018 to the end of the first CHIME/FRB catalog nearly a year later on July 1st, 2019. As discussed in Section~\ref{sec:calibration_challenges} and Section~\ref{sec:automated_calibration_pipeline}, the fluences calculated from this pipeline are \textit{lower limits}, which could be underestimated by a factor of ${\sim}10$ or more for sidelobe detections (see, e.g., Figure~\ref{fig:header_loc_sampling_freq}). As a result, the scientific output gleaned from these fluences has been limited. Note that, because of the limitations described in this paper, as well as a lack of calibration pipeline integration with the injections system \citep{injections_paper}, CHIME/FRB Catalog 1 analyses related to the cumulative FRB fluence distribution \citep{catalog1} and the underlying energy and distance distribution \citep{shin23} used the detection $S/N$ as a proxy for event strength, rather than the fluences from the calibration pipeline.

The limitations of the current flux calibration pipeline are symptomatic of the lag between the immediate scientific results afforded by early FRB detections, and our developing understanding of the technical intricacies involved with CHIME's novel design. Groundbreaking discoveries, like the existence of FRB emission down to $400$\,MHz \citep{abb+19a} and the identification of $18$ new repeating FRB sources \citep{abb+19c,fab+20}, necessitated fluence constraints before subarcminute localization methods and a model of the primary beam had been fully developed.

A future iteration of CHIME/FRB flux calibration, bolstered with an accurate primary beam model and arcminute to sub-arcminute localizations from baseband data \citep{baseband_paper}, could overcome many of the limitations of the first-pass pipeline described in this paper. Now that the CHIME/FRB baseband pipeline was been commissioned \citep{baseband_paper}, the clearest path forward would be to flux-calibrate CHIME/FRB baseband data. In addition to enabling more precise localizations of FRBs within the beam pattern, CHIME/FRB baseband data are also not impacted by the rapid ``clamping'' variations present in the FFT-beamformed intensity data. Since upstream $N^2$ data are both amplitude- and phase-calibrated, CHIME/FRB baseband data are also flux calibrated up to the primary beam, which can be corrected for using our data-driven primary beam model (Section~\ref{sec:beammodel}). This sort of calibration would lead to significant improvement in our burst fluence measurements, as we would be capable of calculating actual accurate fluence estimates rather than lower limits. As such, looking to the future, the next iteration of the CHIME/FRB flux calibration pipeline still holds immense potential for robustly probing fundamental questions related to FRB brightnesses.

\acknowledgements

We give special recognition to CHIME/FRB member Alexander Josephy, who developed the ``header localization'' method mentioned in Section~\ref{sec:localization_limitations} and Figure~\ref{fig:header_loc_sampling_freq}, and who significantly contributed to the development of the CHIME/FRB synthesized beam model. We also thank Ziggy Pleunis for his useful comments during the drafting of this paper.

We acknowledge that CHIME is located on the traditional, ancestral, and unceded territory of the Syilx/Okanagan people. We are grateful to the staff of the Dominion Radio Astrophysical Observatory, which is operated by the National Research Council of Canada. CHIME is funded by a grant from the Canada Foundation for Innovation (CFI) 2012 Leading Edge Fund (Project 31170) and by contributions from the provinces of British Columbia, Qu\'{e}bec and Ontario. The CHIME/FRB Project is funded by a grant from the CFI 2015 Innovation Fund (Project 33213) and by contributions from the provinces of British Columbia and Qu\'{e}bec, and by the Dunlap Institute for Astronomy and Astrophysics at the University of Toronto. Additional support was provided by the Canadian Institute for Advanced Research (CIFAR), McGill University and the McGill Space Institute, the Trottier Family Foundation, and the University of British Columbia. 

\allacks

\bibliographystyle{aasjournal}
\bibliography{references}

\end{document}